\documentclass[usenames,dvipsnames]{aa}
\usepackage{graphicx}
\graphicspath{{./figures/}}
\usepackage[varg]{txfonts}
\usepackage{natbib}
\bibpunct{(}{)}{;}{a}{}{,}
\usepackage[dvipsnames,table,xcdraw]{xcolor}
\usepackage{siunitx}
\usepackage[export]{adjustbox}
\usepackage{subcaption}
\usepackage{pdflscape}
\usepackage{supertabular}
\usepackage{rotating}
\usepackage{ulem}
\usepackage{multirow}
\usepackage{amsmath}
\usepackage[all]{nowidow}

\usepackage[breaklinks=true]{hyperref}
\hypersetup{colorlinks=true, linkcolor=Maroon,
    citecolor=PineGreen,
    bookmarksnumbered=true, pdfborder={0 0 0},unicode,breaklinks}

\defcitealias{limongi2018a}{LC18}
\defcitealias{nomoto2013a}{N13}

\begin{document}

\title{Metallicity-dependent nucleosynthetic yields of Type Ia supernovae
    originating from double detonations of sub-M$_{\text{Ch}}$ white
    dwarfs}

\author{Sabrina~Gronow\inst{1,2}\fnmsep\thanks{\email{sabrinagronow2@gmail.com},
Fellow of the International Max Planck Research School for Astronomy and Cosmic
Physics at Heidelberg University (IMPRS-HD)} \and
Benoit C\^ot\'e\inst{3,4,5} \and
Florian Lach\inst{1,2} \and
Ivo R. Seitenzahl\inst{6} \and
Christine~E.~Collins\inst{7,8} \and
Stuart~A.~Sim\inst{7} \and
Friedrich~K.~R\"{o}pke\inst{2,9} }

\titlerunning{Metallicity-dependent nucleosynthesis yields of SNe\,Ia}
\authorrunning{Gronow et al.}

\institute{%
Zentrum f\"ur Astronomie der Universit\"at Heidelberg,
Astronomisches Rechen-Institut, M\"{o}nchhofstr. 12-14, 69120 Heidelberg, Germany
\and
Heidelberger Institut f\"{u}r Theoretische Studien,
Schloss-Wolfsbrunnenweg 35, 69118 Heidelberg, Germany
\and
Konkoly Observatory, Research Centre for Astronomy and Earth Sciences, E\"otv\"os
Lor\'and Research Network (ELKH), Konkoly Thege Mikl\'os \'ut 15-17, H-1121 Budapest,
Hungary
\and
ELTE E\"otv\"os Lor\'and University, Institute of Physics, Budapest, 1117,
P\'azm\'any Péter S\'et\'any 1/A, Hungary
\and
Joint Institute for Nuclear Astrophysics - Center for the Evolution of the
Elements, USA
\and
School of Science, University of New South Wales, Canberra, ACT 2600, Australia
\and
Astrophysics Research Center, School of Mathematics and Physics, Queen's
University Belfast, Belfast BT7 1NN, Northern Ireland, UK
\and
GSI Helmholtzzentrum f\"{u}r Schwerionenforschung, Planckstraße 1, 64291 Darmstadt, Germany
\and
Zentrum f\"ur Astronomie der Universit\"at Heidelberg, Institut f\"ur
theoretische Astrophysik, Philosophenweg 12, 69120 Heidelberg, Germany}

\date{... ; ...} 

\abstract
{
    Double detonations in sub-Chandrasekhar mass carbon-oxygen white dwarfs
    with helium shell are a potential explosion mechanism for a Type Ia
    supernova. It comprises a shell detonation and subsequent core detonation.
    The focus of our study is on the effect of the progenitor metallicity on
    the nucleosynthetic yields. For this, we compute and analyse a set of
    eleven different models with varying core and shell masses at four
    different metallicities each.  This results in a total of 44 models at
    metallicities between $0.01$\,$Z_\odot$ and $3$\,$Z_\odot$. Our models show
    a strong impact of the metallicity in the high density regime. The presence
    of $^{22}$Ne causes a neutron-excess which shifts the production from
    $^{56}$Ni to stable isotopes such as $^{54}$Fe and $^{58}$Ni in the
    $\alpha$-rich freeze-out regime. The isotopes of the metallicity
    implementation further serve as seed nuclei for additional reactions in the
    shell detonation. Most significantly, the production of $^{55}$Mn increases
    with metallicity confirming the results of previous work. A comparison of
    elemental ratios relative to iron shows a relatively good match to solar
    values for some models. Super-solar values are reached for Mn at
    $3$\,$Z_\odot$ and solar values in some models at \,$Z_\odot$. This
    indicates that the required contribution of Type Ia supernovae originating
    from Chandrasekhar mass WDs can be lower than estimated in previous work to
    reach solar values of [Mn/Fe] at [Fe/H]$=0$.  Our galactic chemical
    evolution models suggest that Type Ia supernovae from sub-Chandrasekhar
    mass white dwarfs, along with core-collapse supernovae, could account for
    more than 80\% of the solar Mn abundance. Using metallicity-dependent Type
    Ia supernova yields helps to reproduce the upward trend of [Mn/Fe] as a
    function of metallicity for the solar neighborhood. These chemical
    evolution predictions, however, depend on the massive star yields adopted
    in the calculations.
}

\maketitle

\section{Introduction}
\label{sec:introduction}
The nucleosynthetic yields from thermonuclear explosions of
carbon-oxygen white dwarfs (CO WDs) depend sensitively on the WD mass and
explosion mechanism. Studies on these were carried out by \citet{iwamoto1999a,
brachwitz2000a, leung2018a, bravo2019a, bravo2019b, leung2020a, lach2020a}
and \citet{gronow2021a} among others. Here, we focus on a further
parameter, namely the metallicity of the WD, which originates from the
metallicity of the zero-age main sequence progenitor star and leads to a
neutron-excess in the WD material.  An enhancement of neutrons can be
achieved in different ways depending on the total mass of the WD. In
Chandrasekhar mass (M$_{\text{Ch}}$) WDs it originates from the initial
metallicity, and, most importantly, from electron capture in high-density
regions. This second effect is not found in sub-M$_{\text{Ch}}$ WDs as high
enough densities are not reached. Instead the metallicity leads to an
initial neutron-excess still present when entering the $\alpha$-rich
freeze-out regime.  The decrease in the electron fraction $Y_e$ increases
the production of neutron-rich iron group elements (IGEs)
\citep{thielemann1986a}.

Nuclear statistical equilibrium (NSE) is not reached in the He detonation of a
sub-M$_{\text{Ch}}$ CO WD with He shell and heavy elements are produced via
$\alpha$-captures. In this density regime the nucleosynthetic yields are
affected by the available seed nuclei of the metallicity implementation
\citep{hashimoto1983a, khokhlov1985a, lach2020a}.  A neutron-excess is
present due to the metallicity and supports the formation of neutron-rich
isotopes.

A metallicity-dependence of the nucleosynthetic yields is expected from
observations.  \citet{hoeflich1998a} find that small changes in the spectra can
be attributed to the metallicity of the WD. They also find that the effect on
the color is stronger and depends on redshift. This is the case as the spectrum
is transferred to different color bands at varying redshifts.  Other effects
are investigated by \citet{umeda1999b} and \citet{iwamoto1999a} who find that
a change in metallicity (and therefore a change in $Y_e$) influences the amount
of $^{54}$Fe, $^{56}$Ni, and $^{58}$Ni produced in the explosion. Furthermore,
\citet{timmes2003a} attribute some variations found in observations of peak
luminosities to different metallicities of the exploding WD. This is in
agreement with work by \citet{mazzali2006a} and \citet{bravo2010a}.
\citet{mazzali2006a} investigate the ($^{54}$Fe\,$+$\,$^{58}$Ni) / $^{56}$Ni
ratio as a source of the scatter in the light curves and find that their
bolometric light curves are fainter with higher ratios (differences of
0.25\,mag). Their values for the luminosity decrease over 15\,days after peak,
$\Delta m_{15}$, compare well with observations.  In addition,
\citet{timmes2003a} derive a linear relation between the metallicity and
$^{56}$Ni mass produced in the explosion. It is apparent in their models of
$\mathrm{M}_\mathrm{Ch}$ WDs at $1/3$\,$Z_\odot$ to $3$\,$Z_\odot$ that the amount of
produced $^{56}$Ni decreases by 25\% going to higher metallicity. A
determination of the metallicity from SN observations is attempted by
\citet{lentz2000a} and \citet{taubenberger2008a}, though they are affected by
large uncertainties originating from the imprecision of the observations
themselves as well as potential degeneracies between metallicity and other
effects on the spectra.

SNe~Ia play a key role in galactic chemical evolution (GCE)
\citep[e.g.,][]{greggio1983a, matteucci1986a, kobayashi1998a, lach2020a}. The
nucleosynthetic yields which are found in explosion simulations are crucial
input parameters for GCE models.  Comparing GCE predictions with the
composition of the Milky Way and other galaxies helps to constrain the
explosion mechanism and identify the importance of various proposed SN~Ia
progenitors. Previous GCE studies suggested that multiple SN~Ia channels should
have contributed to the synthesis of Mn (e.g., \citealt{seitenzahl2013b,
cescutti2017d}), including a possible $\sim25-75$\,\% contribution from
sub-M$_\mathrm{Ch}$ explosions in the solar neighbourhood (e.g.,
\citealt{seitenzahl2013b, kobayashi2019b, eitner2020a}).  Additional GCE
studies suggested that chemical evolution trends derived from spectroscopy are
better reproduced with models that include metallicity-dependent SN~Ia yields
(e.g., \citealt{kobayashi1998a,cescutti2008a,north2012a}). Work by
\citet{de2020a} suggests that the dominant progenitors are sub-M$_\mathrm{Ch}$
WDs at early times shifting to M$_\mathrm{Ch}$ WDs at later times based on
their comparison of data from dwarf spheroidal galaxies with extended star
formation history to theoretical models.

We investigate the dependence of the nucleosynthetic yields of
thermonuclear explosions of sub-M$_{\text{Ch}}$ CO WDs with He shell on the
metallicity of the progenitor star as a follow-up study to the work presented
in \citet{gronow2021a}. The simulations of \citet{gronow2021a} follow a double
detonation of the WD. In this scenario a He detonation is ignited at the base
of the shell due to thermal instabilities. It causes a second, core detonation
when the shock wave converges off-center in the core (converging shock
mechanism), or when the He detonation wave converges at the antipode to its
ignition spot (scissors mechanism). In a third scenario, a carbon detonation is
ignited at the core-shell interface directly after He ignition (edge-lit
mechanism). \citet{gronow2021a} compute a set of models exploring different
core and He shell masses at solar metallicity. On the basis of these models
three more metallicity values of the WD are examined in this work spanning
the range from $0.01$\,$Z_\odot$ to $3$\,$Z_\odot$. This work is one of the
first to cover the parameter space for sub-M$_{\text{Ch}}$ CO WDs with He
shell \citep[see also][]{leung2020a}.  Other work \citep[e.g.,][]{sim2010a,
shen2018b} investigates pure detonations of sub-M$_{\text{Ch}}$ CO WDs.

The methods and model setup are described in Section \ref{sec:models_methods}.
The results of the nucleosynthesis calculations are presented in Section
\ref{sec:results} including a comparison of elemental ratios relative to iron
to solar values. In Section \ref{sec:discussion} the results are compared to
previous works. We discuss the implication of our SN~Ia yields in a GCE context
in Section~\ref{sec:gce}. Conclusions are drawn in Section \ref{sec:summary}.

\section{Methods and Model setup}
\label{sec:models_methods}
In our hydrodynamic simulations of explosions of CO WDs with He shells, the
metallicity is represented by $^{14}$N and $^{22}$Ne. During the evolution of
the WD progenitor star and its companion $^{14}$N is formed in hydrogen burning
via the CNO cycle from pre-existing C, N, and O at zero-age main sequence. In
subsequent He burning of the WD progenitor, it is converted to $^{22}$Ne via
$^{14}$N($\alpha$,$\gamma$)$^{18}$F($\beta^+$,
$\nu_e$)$^{18}$O($\alpha$,$\gamma$)$^{22}$Ne. There, the metallicity of the
progenitor star ends up as $^{14}$N in the He shell and $^{22}$Ne in the CO
core. A homogeneous distribution of $^{22}$Ne is employed in the core.  This is
a reasonable assumption as sedimentation of $^{22}$Ne in the rather extended
sub-M$_{\text{Ch}}$ WDs is not expected to have a significant effect
\citep[see][]{bildsten2001a, deloye2002a, garcia-berro2008a}.

In a postprocessing step, we increased the number of species included to
represent the metallicity in order to obtain detailed nucleosynthetic yields.
In our simulations the isotopes coming from the implementation of the
metallicity influence the final yields.

Our study is based on results of the hydrodynamics simulations of
\citet{gronow2021a}. Their explosion simulations were carried out with the
\textsc{Arepo} code \citep{springel2010a} in 3D.  Details on the method can be
found in \citet{gronow2021a} and references therein. The models cover a range
of different core and shell masses. The core mass of their models is between
$0.8$\,$M_\odot$ and $1.1$\,$M_\odot$ with shell masses of $0.02$\,$M_\odot$ to
$0.1$\,$M_\odot$.

\begin{figure}
    \centering
    \includegraphics[width=0.48\textwidth]{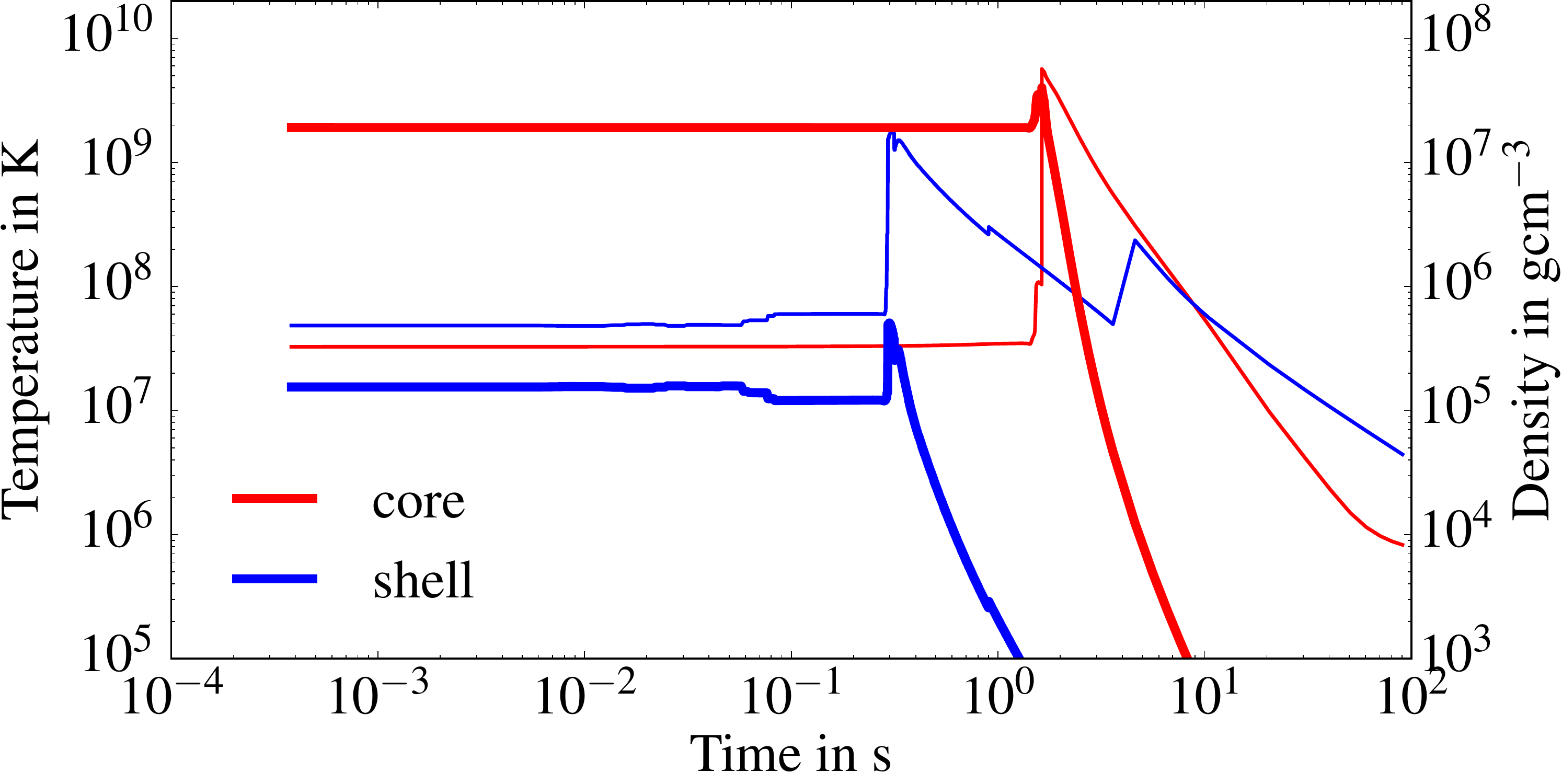}
	\caption{Temperature (thin lines) and density (thick lines)
	evolution of a shell (blue) and core (red) tracer particle of Model
	M10\_03\_1.}
\label{fig:tracer_evol}
\end{figure}

We used the temperature and density evolution of the two million tracer
particles from their explosion simulations \citep{gronow2021a}. These tracers
had a random distribution in the initial WD which sampled the mass distribution
of the WD and represented equal fractions of the material. They allow us to
determine detailed nucleosynthetic yields and the ejecta structure of the SN~Ia
explosion models \citep{travaglio2004a}. The evolution of the
temperature and density of tracers which were initially located in the He shell
and CO core of Model M10\_03\_1 are shown in Figure~\ref{fig:tracer_evol} in
red and blue, respectively. The tracers were chosen at random while,
nevertheless, representing a typical evolution of tracers in these regions. The
peaks in the profiles illustrate well that the He detonation reached the tracer
particle in the He shell in the first $0.4$\,s before the C detonation had an
impact on the tracer particle in the core.  Following this, the tracer particle
from the He shell exhibits another rise in temperature and density as a shock
wave originating from the C detonation reaches its position.
\citet{gronow2021a} carried out a postprocessing step involving a large nuclear
reaction network consisting of 384 isotopes. The models in their study assumed
solar metallicity of the zero-age main sequence progenitor star.

We followed up on this by re-calculating the postprocessing step using the
parameters of the hydrodynamic simulations which are listed in Tables 1 and 2
of \citet{gronow2021a}, and \citet{gronow2020a} for Models M2a and M2a\_pp. Our
additional models assume three different metallicities, $0.01$, $0.1$, and
$3$\,$Z_\odot$. For this, we scaled the solar abundances given by
\citet{asplund2009a}. However, according to \citet{prantzos2018a} at sub-solar
metallicity we kept the abundance ratios of the $\alpha$-elements fixed (i.e.,
[C/Fe]=0.18, [O/Fe]=0.47, [Mg/Fe]=0.27, [Si/Fe]=0.37, [S/Fe]=0.35,
[Ar/Fe]=0.35, [Ca/Fe]=0.33, [Ti/Fe]=0.23) which is motivated by observations of
low-metallicity stars. Similar to \citet{gronow2021a}, we used this set of
elements to represent metallicity in the postprocessing step along with the
initial profile of the composition.  In the initial profile of the hydrodynamic
simulations, the metallicity was set by including $^{14}$N and $^{22}$Ne in the
shell and core, respectively \citep[see][]{gronow2021a}. We point out that some
core material was mixed into the shell during the relaxation step (see
\citealp{gronow2020a} and \citealp{gronow2021a} for a detailed description of
the relaxation step) adding some $^{12}$C and $^{22}$Ne to the shell. A total
of 384 species was followed in the postprocessing step. In the same way as
\citet{gronow2021a} and following the methods in \citet{pakmor2012b}, the
REACLIB data base \citep{rauscher2000a} in its 2014 version was used. Weak
reaction rates were taken from \citet{langanke2001a}.

Major differences between our approach and a full re-calculation of the
hydrodynamics were not expected. This was the case since the changes in the
$^{14}$N and $^{22}$Ne abundances at the different metallicities do not alter
the energy release in the hydrodynamic simulations significantly. It is
different to deflagrations where the buoyancy, and therefore Rayleigh-Taylor
instabilities, depend on $Y_e$. In contrast to detonations, the propagation of
a deflagration front is thus affected by the metallicity
\citep[e.g.,][]{meakin2009a}.  The assumption we made here is confirmed by the
comparison of the models presented in Table \ref{tab:abund01}. Model M2a is
taken from \citet{gronow2020a}.  The model was calculated at zero metallicity
and has a total mass of $1.05\,M_\odot$ with a He shell of $0.07\,M_\odot$ at
He ignition. Model M10\_05\_1, on the other hand, has a similar mass
configuration, though it was calculated at solar metallicity (Model M10\_05 in
\citealp{gronow2021a}).  Model M2a\_pp is the same model as Model M2a, but the
postprocessing step was calculated with solar metallicity instead of zero
metallicity. An inspection of the abundances of Models M2a\_pp and M10\_05\_1
at $t=100\,\text{s}$ after He detonation ignition shows that the results of the
postprocessing step with varying metallicities are in reasonable good agreement
with a full re-calculation of the hydrodynamic model The maximum difference in
the yields produced in the core detonation is only 10\%, while the maximum
difference is 50\% in the He detonation (excluding $^{12}$C in both). However,
differences in the yields produced in the He detonation can in part be
attributed to the slightly different setups of Model M2a (and therefore Model
M2a\_pp) and Model M10\_05\_1 at the beginning of the relaxation simulation,
with the differences in the total and shell masses being less than 1\% (see
\citealp{gronow2021a} for an explanation of the difference).  Generally, the
contribution of the yields from the He detonation to the total nucleosynthetic
yields are small compared to those of the core detonation. Our approach is thus
sufficient to derive nucleosynthetic yields for GCE studies.  It saves
significant computational costs as additional 3D hydrodynamical simulations of
the explosion do not need to be carried out.  There might, nevertheless, be
slight differences visible in the observables because they are sensitive to the
products of the He shell detonation \citep{hoeflich1996b, nugent1997a,
kromer2010a}.

Our models are named based on the initial core mass (first two digits), the
initial shell mass (two digits) and metallicity relative to solar. Models
M08\_10\_r and M09\_10\_r of \citet{gronow2021a} are renamed to M08\_10 and
M09\_10, respectively, for simplicity. The use of names such as M10\_05 refers
to all models with an initial core mass of $1.0\,M_\odot$ and shell mass of
$0.05\,M_\odot$, therefore combining names of four models at different
metallicity. The nucleosynthesis yields of the models will be uploaded to the
supernova archive HESMA\footnote[1]{\url{https://hesma.h-its.org}}
\citep{kromer2017a}.

\begin{table*}
    \caption{Final abundances for Model M2a$^{(1)}$, M2a\_pp, and M10\_05\_1$^{(2)}$.}
    \label{tab:abund01}
    \centering
\begin{tabular}{@{}lrrrrrr}
        \hline
        & \multicolumn{3}{c}{He detonation} & \multicolumn{3}{c}{core detonation} \\
        & M2a$^{(1)}$ & M2a\_pp & M10\_05\_1$^{(2)}$ & M2a$^{(1)}$ & M2a\_pp & M10\_05\_1$^{(2)}$ \\
        & [M$_\odot$] & [M$_\odot$] & [M$_\odot$] & [M$_\odot$] & [M$_\odot$] & [M$_\odot$] \\ \hline
        $^4$He & $2.3\times10^{-2}$ & $2.2\times10^{-2}$ & $2.0\times10^{-2}$ & $5.0\times10^{-3}$ & $4.2\times10^{-3}$ & $4.6\times10^{-3}$ \\
        $^{12}$C & $1.0\times10^{-4}$ & $1.0\times10^{-4}$ & $4.0\times10^{-5}$ & $8.9\times10^{-4}$ & $8.6\times10^{-4}$ & $4.4\times10^{-4}$ \\
        $^{16}$O & $7.4\times10^{-3}$ & $6.8\times10^{-3}$ & $9.3\times10^{-3}$ & $5.2\times10^{-2}$ & $5.6\times10^{-2}$ & $6.1\times10^{-2}$ \\
        $^{28}$Si & $8.9\times10^{-3}$ & $9.6\times10^{-3}$ & $1.3\times10^{-2}$ & $1.6\times10^{-1}$ & $1.7\times10^{-1}$ & $1.6\times10^{-1}$ \\
        $^{32}$S & $3.2\times10^{-3}$ & $3.6\times10^{-3}$ & $4.9\times10^{-3}$ & $1.1\times10^{-1}$ & $1.0\times10^{-1}$ & $9.6\times10^{-2}$ \\
\hline
    \end{tabular}
    \tablebib{(1)~\citet{gronow2020a}, (2)~\citet{gronow2021a}}
\end{table*}

\begin{table*}
    \caption{Final abundances for Model M10\_05\_001, M10\_05\_01, and M10\_05\_3.}
    \label{tab:abund02}
    \centering
    \begin{tabular}{@{}lrrrrrr}
        \hline
        & \multicolumn{3}{c}{He detonation} & \multicolumn{3}{c}{core detonation} \\
        & M10\_05\_001 & M10\_05\_01 & M10\_05\_3 & M10\_05\_001 & M10\_05\_01 & M10\_05\_3 \\
        & [M$_\odot$] & [M$_\odot$] & [M$_\odot$] & [M$_\odot$] & [M$_\odot$] & [M$_\odot$]\\ \hline
        $^4$He & $2.0\times10^{-2}$ & $2.0\times10^{-2}$ & $1.9\times10^{-2}$ & $5.0\times10^{-3}$ & $5.0\times10^{-3}$ & $3.3\times10^{-3}$ \\
        $^{12}$C & $4.0\times10^{-5}$ & $4.0\times10^{-5}$ & $4.0\times10^{-5}$ & $4.4\times10^{-4}$ & $4.4\times10^{-4}$ & $4.3\times10^{-4}$ \\
        $^{16}$O & $9.2\times10^{-3}$ & $9.2\times10^{-3}$ & $9.7\times10^{-3}$ & $6.0\times10^{-2}$ & $6.0\times10^{-2}$ & $6.1\times10^{-2}$ \\
        $^{28}$Si & $1.3\times10^{-2}$ & $1.3\times10^{-2}$ & $1.3\times10^{-2}$ & $1.6\times10^{-1}$ & $1.6\times10^{-1}$ & $1.6\times10^{-1}$ \\
        $^{32}$S & $5.0\times10^{-3}$ & $5.0\times10^{-3}$ & $4.9\times10^{-3}$ & $9.9\times10^{-2}$ & $9.9\times10^{-2}$ & $8.4\times10^{-2}$ \\
\hline
    \end{tabular}
\end{table*}

\begin{table*}
    \caption{Final abundances for Model M10\_10\_001, M10\_10\_01, M10\_10\_1, and M10\_10\_3.}
    \label{tab:abund03}
    \centering
    \begin{tabular}{@{}lrrrrrrrr}
        \hline
        & \multicolumn{4}{c}{He detonation} & \multicolumn{4}{c}{core detonation} \\
        & M10\_10\_001 & M10\_10\_01 & M10\_10\_1$^{(2)}$ & M10\_10\_3 & M10\_10\_001 & M10\_10\_01 & M10\_10\_1$^{(2)}$ & M10\_10\_3 \\
        & [M$_\odot$] & [M$_\odot$] & [M$_\odot$] & [M$_\odot$] & [M$_\odot$] & [M$_\odot$] & [M$_\odot$] & [M$_\odot$]\\ \hline
        $^4$He & $2.1\times10^{-2}$ & $2.1\times10^{-2}$ & $2.1\times10^{-2}$ & $2.0\times10^{-2}$ & $7.1\times10^{-3}$ & $7.1\times10^{-3}$ & $6.5\times10^{-3}$ & $4.5\times10^{-3}$ \\
        $^{12}$C & $7.4\times10^{-6}$ & $7.3\times10^{-6}$ & $1.1\times10^{-5}$ & $7.5\times10^{-6}$ & $2.5\times10^{-6}$ & $2.5\times10^{-6}$ & $1.7\times10^{-5}$ & $1.1\times10^{-6}$ \\
        $^{16}$O & $3.0\times10^{-3}$ & $3.0\times10^{-3}$ & $3.1\times10^{-3}$ & $3.2\times10^{-3}$ & $2.7\times10^{-3}$ & $2.7\times10^{-3}$ & $2.7\times10^{-3}$ & $2.7\times10^{-3}$ \\
        $^{28}$Si & $3.7\times10^{-2}$ & $3.7\times10^{-2}$ & $3.7\times10^{-2}$ & $3.8\times10^{-2}$ & $7.3\times10^{-2}$ & $7.3\times10^{-2}$ & $7.3\times10^{-2}$ & $7.3\times10^{-2}$ \\
        $^{32}$S & $1.6\times10^{-2}$ & $1.6\times10^{-2}$ & $1.6\times10^{-2}$ & $1.6\times10^{-2}$ & $5.6\times10^{-2}$ & $5.6\times10^{-2}$ & $5.4\times10^{-2}$ & $4.8\times10^{-2}$ \\
\hline
    \end{tabular}
    \tablebib{(2)~\citet{gronow2021a}}
\end{table*}

\begin{table*}
    \caption{Final abundances for Model M10\_03\_001, M10\_03\_01, M10\_03\_1, and M10\_03\_3.}
    \label{tab:abund04}
    \centering
    \begin{tabular}{@{}lrrrrrrrr}
        \hline
        & \multicolumn{4}{c}{He detonation} & \multicolumn{4}{c}{core detonation} \\
        & M10\_03\_001 & M10\_03\_01 & M10\_03\_1$^{(2)}$ & M10\_03\_3 & M10\_03\_001 & M10\_03\_01 & M10\_03\_1$^{(2)}$ & M10\_03\_3 \\
        & [M$_\odot$] & [M$_\odot$] & [M$_\odot$] & [M$_\odot$] & [M$_\odot$] & [M$_\odot$] & [M$_\odot$] & [M$_\odot$]\\ \hline
        $^4$He & $1.3\times10^{-2}$ & $1.3\times10^{-2}$ & $1.3\times10^{-2}$ & $1.2\times10^{-2}$ & $5.5\times10^{-3}$ & $5.5\times10^{-3}$ & $5.1\times10^{-3}$ & $3.6\times10^{-3}$ \\
        $^{12}$C & $7.7\times10^{-4}$ & $7.7\times10^{-4}$ & $7.6\times10^{-4}$ & $7.5\times10^{-4}$ & $1.2\times10^{-3}$ & $1.2\times10^{-3}$ & $1.2\times10^{-3}$ & $1.2\times10^{-3}$\\
        $^{16}$O & $6.6\times10^{-3}$ & $6.6\times10^{-3}$ & $6.8\times10^{-3}$ & $7.1\times10^{-3}$ & $4.8\times10^{-2}$ & $4.8\times10^{-2}$ & $4.9\times10^{-2}$ & $4.9\times10^{-2}$\\
        $^{28}$Si & $8.9\times10^{-3}$ & $8.9\times10^{-3}$ & $8.9\times10^{-3}$ & $9.0\times10^{-3}$ & $1.5\times10^{-1}$ & $1.5\times10^{-1}$ & $1.5\times10^{-1}$ & $1.5\times10^{-1}$\\
        $^{32}$S & $3.7\times10^{-3}$ & $3.7\times10^{-3}$ & $3.7\times10^{-3}$ & $3.7\times10^{-3}$ & $9.4\times10^{-2}$ & $9.4\times10^{-2}$ & $9.1\times10^{-2}$ & $8.0\times10^{-2}$\\
\hline
    \end{tabular}
    \tablebib{(2)~\citet{gronow2021a}}
\end{table*}

\begin{table*}
    \caption{Final abundances for Model M10\_02\_001, M10\_02\_01, M10\_02\_1, and M10\_02\_3.}
    \label{tab:abund05}
    \centering
    \begin{tabular}{@{}lrrrrrrrr}
        \hline
        & \multicolumn{4}{c}{He detonation} & \multicolumn{4}{c}{core detonation} \\
        & M10\_02\_001 & M10\_02\_01 & M10\_02\_1$^{(2)}$ & M10\_02\_3 & M10\_02\_001 & M10\_02\_01 & M10\_02\_1$^{(2)}$ & M10\_02\_3 \\
        & [M$_\odot$] & [M$_\odot$] & [M$_\odot$] & [M$_\odot$] & [M$_\odot$] & [M$_\odot$] & [M$_\odot$] & [M$_\odot$]\\ \hline
        $^4$He & $1.4\times10^{-2}$ & $1.4\times10^{-2}$ & $1.3\times10^{-2}$ & $1.3\times10^{-2}$ & $4.2\times10^{-3}$ & $4.2\times10^{-3}$ & $3.8\times10^{-3}$ & $2.7\times10^{-3}$ \\
        $^{12}$C & $1.7\times10^{-3}$ & $1.7\times10^{-3}$ & $1.7\times10^{-3}$ & $1.6\times10^{-3}$ & $2.0\times10^{-3}$ & $2.0\times10^{-3}$ & $1.9\times10^{-3}$ & $1.9\times10^{-3}$\\
        $^{16}$O & $1.8\times10^{-3}$ & $1.8\times10^{-3}$ & $1.9\times10^{-3}$ & $2.0\times10^{-3}$ & $5.7\times10^{-2}$ & $5.7\times10^{-2}$ & $5.7\times10^{-2}$ & $5.8\times10^{-2}$\\
        $^{28}$Si & $2.9\times10^{-3}$ & $2.9\times10^{-3}$ & $2.9\times10^{-3}$ & $3.0\times10^{-3}$ & $1.7\times10^{-1}$ & $1.7\times10^{-1}$ & $1.7\times10^{-1}$ & $1.7\times10^{-1}$\\
        $^{32}$S & $1.6\times10^{-3}$ & $1.6\times10^{-3}$ & $1.6\times10^{-3}$ & $1.7\times10^{-3}$ & $1.1\times10^{-1}$ & $1.1\times10^{-1}$ & $1.0\times10^{-1}$ & $9.0\times10^{-2}$\\
\hline
    \end{tabular}
    \tablebib{(2)~\citet{gronow2021a}}
\end{table*}

\begin{table*}
    \caption{Final abundances for Model M09\_10\_001, M09\_10\_01, M09\_10\_1, and M09\_10\_3.}
    \label{tab:abund06}
    \centering
    \begin{tabular}{@{}lrrrrrrrr}
        \hline
        & \multicolumn{4}{c}{He detonation} & \multicolumn{4}{c}{core detonation} \\
        & M09\_10\_001 & M09\_10\_01 & M09\_10\_1$^{(2)}$ & M09\_10\_3 & M09\_10\_001 & M09\_10\_01 & M09\_10\_1$^{(2)}$ & M09\_10\_3 \\
        & [M$_\odot$] & [M$_\odot$] & [M$_\odot$] & [M$_\odot$] & [M$_\odot$] & [M$_\odot$] & [M$_\odot$] & [M$_\odot$]\\ \hline
        $^4$He & $3.2\times10^{-2}$ & $3.2\times10^{-2}$ & $3.2\times10^{-2}$ & $3.1\times10^{-2}$ & $4.2\times10^{-3}$ & $4.2\times10^{-3}$ & $3.9\times10^{-3}$ & $2.8\times10^{-3}$ \\
        $^{12}$C & $3.8\times10^{-5}$ & $3.8\times10^{-5}$ & $3.9\times10^{-5}$ & $3.8\times10^{-5}$ & $1.4\times10^{-4}$ & $1.3\times10^{-4}$ & $1.3\times10^{-4}$ & $1.3\times10^{-4}$\\
        $^{16}$O & $8.4\times10^{-3}$ & $8.4\times10^{-3}$ & $8.5\times10^{-3}$ & $8.8\times10^{-3}$ & $5.5\times10^{-2}$ & $5.5\times10^{-2}$ & $5.5\times10^{-2}$ & $5.6\times10^{-2}$\\
        $^{28}$Si & $1.3\times10^{-2}$ & $1.3\times10^{-2}$ & $1.3\times10^{-2}$ & $1.3\times10^{-2}$ & $1.5\times10^{-1}$ & $1.5\times10^{-1}$ & $1.6\times10^{-1}$ & $1.5\times10^{-1}$\\
        $^{32}$S & $4.4\times10^{-3}$ & $4.3\times10^{-3}$ & $4.3\times10^{-3}$ & $4.3\times10^{-3}$ & $9.5\times10^{-2}$ & $9.5\times10^{-2}$ & $9.2\times10^{-2}$ & $8.1\times10^{-2}$\\
\hline
    \end{tabular}
    \tablebib{(2)~\citet{gronow2021a}}
\end{table*}

\begin{table*}
    \caption{Final abundances for Model M09\_05\_001, M09\_05\_01, M09\_05\_1, and M09\_05\_3.}
    \label{tab:abund07}
    \centering
    \begin{tabular}{@{}lrrrrrrrr}
        \hline
        & \multicolumn{4}{c}{He detonation} & \multicolumn{4}{c}{core detonation} \\
        & M09\_05\_001 & M09\_05\_01 & M09\_05\_1$^{(2)}$ & M09\_05\_3 & M09\_05\_001 & M09\_05\_01 & M09\_05\_1$^{(2)}$ & M09\_05\_3 \\
        & [M$_\odot$] & [M$_\odot$] & [M$_\odot$] & [M$_\odot$] & [M$_\odot$] & [M$_\odot$] & [M$_\odot$] & [M$_\odot$]\\ \hline
        $^4$He & $2.6\times10^{-2}$ & $2.6\times10^{-2}$ & $2.5\times10^{-2}$ & $2.5\times10^{-2}$ & $2.1\times10^{-3}$ & $2.1\times10^{-3}$ & $1.9\times10^{-3}$ & $1.3\times10^{-3}$ \\
        $^{12}$C & $4.4\times10^{-4}$ & $4.4\times10^{-4}$ & $4.3\times10^{-4}$ & $4.3\times10^{-4}$ & $2.7\times10^{-3}$ & $2.7\times10^{-3}$ & $2.6\times10^{-3}$ & $2.6\times10^{-3}$\\
        $^{16}$O & $7.1\times10^{-3}$ & $7.1\times10^{-3}$ & $7.3\times10^{-3}$ & $7.6\times10^{-3}$ & $7.7\times10^{-2}$ & $7.7\times10^{-2}$ & $7.8\times10^{-2}$ & $7.9\times10^{-2}$\\
        $^{28}$Si & $1.0\times10^{-2}$ & $1.0\times10^{-2}$ & $1.0\times10^{-2}$ & $1.0\times10^{-2}$ & $1.9\times10^{-1}$ & $1.9\times10^{-1}$ & $1.9\times10^{-1}$ & $1.9\times10^{-1}$\\
        $^{32}$S & $4.5\times10^{-3}$ & $4.5\times10^{-3}$ & $4.4\times10^{-3}$ & $4.5\times10^{-3}$ & $1.1\times10^{-1}$ & $1.1\times10^{-1}$ & $1.1\times10^{-1}$ & $9.7\times10^{-2}$\\
\hline
    \end{tabular}
    \tablebib{(2)~\citet{gronow2021a}}
\end{table*}

\begin{table*}
    \caption{Final abundances for Model M09\_03\_001, M09\_03\_01, M09\_03\_1, and M09\_03\_3.}
    \label{tab:abund08}
    \centering
    \begin{tabular}{@{}lrrrrrrrr}
        \hline
        & \multicolumn{4}{c}{He detonation} & \multicolumn{4}{c}{core detonation} \\
        & M09\_03\_001 & M09\_03\_01 & M09\_03\_1$^{(2)}$ & M09\_03\_3 & M09\_03\_001 & M09\_03\_01 & M09\_03\_1$^{(2)}$ & M09\_03\_3 \\
        & [M$_\odot$] & [M$_\odot$] & [M$_\odot$] & [M$_\odot$] & [M$_\odot$] & [M$_\odot$] & [M$_\odot$] & [M$_\odot$]\\ \hline
        $^4$He & $1.5\times10^{-2}$ & $1.5\times10^{-2}$ & $1.5\times10^{-2}$ & $1.4\times10^{-2}$ & $6.7\times10^{-4}$ & $6.8\times10^{-4}$ & $5.8\times10^{-4}$ & $3.3\times10^{-4}$ \\
        $^{12}$C & $3.5\times10^{-3}$ & $3.5\times10^{-3}$ & $3.5\times10^{-3}$ & $3.4\times10^{-3}$ & $4.9\times10^{-3}$ & $4.9\times10^{-3}$ & $4.9\times10^{-3}$ & $4.7\times10^{-3}$\\
        $^{16}$O & $3.8\times10^{-3}$ & $3.8\times10^{-3}$ & $3.9\times10^{-3}$ & $4.2\times10^{-3}$ & $9.2\times10^{-2}$ & $9.2\times10^{-2}$ & $9.2\times10^{-2}$ & $9.3\times10^{-2}$\\
        $^{28}$Si & $5.8\times10^{-3}$ & $5.8\times10^{-3}$ & $5.8\times10^{-3}$ & $5.9\times10^{-3}$ & $2.2\times10^{-1}$ & $2.2\times10^{-1}$ & $2.2\times10^{-1}$ & $2.2\times10^{-1}$\\
        $^{32}$S & $2.7\times10^{-3}$ & $2.7\times10^{-3}$ & $2.8\times10^{-3}$ & $2.9\times10^{-3}$ & $1.3\times10^{-1}$ & $1.3\times10^{-1}$ & $1.3\times10^{-1}$ & $1.1\times10^{-1}$\\
\hline
    \end{tabular}
    \tablebib{(2)~\citet{gronow2021a}}
\end{table*}

\begin{table*}
    \caption{Final abundances for Model M08\_10\_001, M08\_10\_01, M08\_10\_1, and M08\_10\_3.}
    \label{tab:abund09}
    \centering
    \begin{tabular}{@{}lrrrrrrrr}
        \hline
        & \multicolumn{4}{c}{He detonation} & \multicolumn{4}{c}{core detonation} \\
        & M08\_10\_001 & M08\_10\_01 & M08\_10\_1$^{(2)}$ & M08\_10\_3 & M08\_10\_001 & M08\_10\_01 & M08\_10\_1$^{(2)}$ & M08\_10\_3 \\
        & [M$_\odot$] & [M$_\odot$] & [M$_\odot$] & [M$_\odot$] & [M$_\odot$] & [M$_\odot$] & [M$_\odot$] & [M$_\odot$]\\ \hline
        $^4$He & $3.6\times10^{-2}$ & $3.6\times10^{-2}$ & $3.6\times10^{-2}$ & $3.4\times10^{-2}$ & $1.5\times10^{-3}$ & $1.5\times10^{-3}$ & $1.4\times10^{-3}$ & $9.9\times10^{-4}$ \\
        $^{12}$C & $1.2\times10^{-4}$ & $1.2\times10^{-4}$ & $1.2\times10^{-4}$ & $1.2\times10^{-4}$ & $1.1\times10^{-3}$ & $1.1\times10^{-3}$ & $1.1\times10^{-3}$ & $1.0\times10^{-3}$\\
        $^{16}$O & $9.1\times10^{-3}$ & $9.1\times10^{-3}$ & $9.3\times10^{-3}$ & $9.6\times10^{-3}$ & $8.0\times10^{-2}$ & $8.0\times10^{-2}$ & $8.1\times10^{-2}$ & $8.2\times10^{-2}$\\
        $^{28}$Si & $1.3\times10^{-2}$ & $1.2\times10^{-2}$ & $1.3\times10^{-2}$ & $1.3\times10^{-2}$ & $1.9\times10^{-1}$ & $1.9\times10^{-1}$ & $1.9\times10^{-1}$ & $1.9\times10^{-1}$\\
        $^{32}$S & $5.6\times10^{-3}$ & $5.6\times10^{-3}$ & $5.5\times10^{-3}$ & $5.6\times10^{-3}$ & $1.1\times10^{-1}$ & $1.1\times10^{-1}$ & $1.1\times10^{-1}$ & $9.6\times10^{-2}$\\
\hline
    \end{tabular}
    \tablebib{(2)~\citet{gronow2021a}}
\end{table*}

\begin{table*}
    \caption{Final abundances for Model M08\_05\_001, M08\_05\_01, M08\_05\_1, and M08\_05\_3.}
    \label{tab:abund10}
    \centering
    \begin{tabular}{@{}lrrrrrrrr}
        \hline
        & \multicolumn{4}{c}{He detonation} & \multicolumn{4}{c}{core detonation} \\
        & M08\_05\_001 & M08\_05\_01 & M08\_05\_1$^{(2)}$ & M08\_05\_3 & M08\_05\_001 & M08\_05\_01 & M08\_05\_1$^{(2)}$ & M08\_05\_3 \\
        & [M$_\odot$] & [M$_\odot$] & [M$_\odot$] & [M$_\odot$] & [M$_\odot$] & [M$_\odot$] & [M$_\odot$] & [M$_\odot$]\\ \hline
        $^4$He & $2.8\times10^{-2}$ & $2.8\times10^{-2}$ & $2.7\times10^{-2}$ & $2.6\times10^{-2}$ & $9.8\times10^{-5}$ & $9.8\times10^{-5}$ & $8.2\times10^{-5}$ & $4.6\times10^{-5}$ \\
        $^{12}$C & $2.3\times10^{-3}$ & $2.3\times10^{-3}$ & $2.3\times10^{-3}$ & $2.3\times10^{-3}$ & $7.5\times10^{-3}$ & $7.5\times10^{-3}$ & $7.5\times10^{-3}$ & $7.2\times10^{-3}$\\
        $^{16}$O & $6.1\times10^{-3}$ & $6.1\times10^{-3}$ & $6.3\times10^{-3}$ & $6.6\times10^{-3}$ & $1.2\times10^{-1}$ & $1.2\times10^{-1}$ & $1.2\times10^{-1}$ & $1.2\times10^{-1}$\\
        $^{28}$Si & $9.2\times10^{-3}$ & $9.2\times10^{-3}$ & $9.3\times10^{-3}$ & $9.4\times10^{-3}$ & $2.3\times10^{-1}$ & $2.3\times10^{-1}$ & $2.3\times10^{-1}$ & $2.3\times10^{-1}$\\
        $^{32}$S & $4.7\times10^{-3}$ & $4.7\times10^{-3}$ & $4.7\times10^{-3}$ & $4.8\times10^{-3}$ & $1.3\times10^{-1}$ & $1.3\times10^{-1}$ & $1.3\times10^{-1}$ & $1.1\times10^{-1}$\\
\hline
    \end{tabular}
    \tablebib{(2)~\citet{gronow2021a}}
\end{table*}

\begin{table*}
    \caption{Final abundances for Model M08\_03\_001, M08\_03\_01, M08\_03\_1, and M08\_03\_3.}
    \label{tab:abund11}
    \centering
    \begin{tabular}{@{}lrrrrrrrr}
        \hline
        & \multicolumn{4}{c}{He detonation} & \multicolumn{4}{c}{core detonation} \\
        & M08\_03\_001 & M08\_03\_01 & M08\_03\_1$^{(2)}$ & M08\_03\_3 & M08\_03\_001 & M08\_03\_01 & M08\_03\_1$^{(2)}$ & M08\_03\_3 \\
        & [M$_\odot$] & [M$_\odot$] & [M$_\odot$] & [M$_\odot$] & [M$_\odot$] & [M$_\odot$] & [M$_\odot$] & [M$_\odot$]\\ \hline
        $^4$He & $1.9\times10^{-2}$ & $1.9\times10^{-2}$ & $1.8\times10^{-2}$ & $1.8\times10^{-2}$ & $3.5\times10^{-6}$ & $3.5\times10^{-6}$ & $3.2\times10^{-6}$ & $2.5\times10^{-6}$ \\
        $^{12}$C & $3.4\times10^{-3}$ & $3.4\times10^{-3}$ & $3.3\times10^{-3}$ & $3.3\times10^{-3}$ & $1.3\times10^{-2}$ & $1.3\times10^{-2}$ & $1.2\times10^{-2}$ & $1.2\times10^{-2}$\\
        $^{16}$O & $2.5\times10^{-3}$ & $2.5\times10^{-3}$ & $2.6\times10^{-3}$ & $2.8\times10^{-3}$ & $1.4\times10^{-1}$ & $1.4\times10^{-1}$ & $1.4\times10^{-1}$ & $1.5\times10^{-1}$\\
        $^{28}$Si & $4.2\times10^{-3}$ & $4.2\times10^{-3}$ & $4.2\times10^{-3}$ & $4.4\times10^{-3}$ & $2.6\times10^{-1}$ & $2.6\times10^{-1}$ & $2.6\times10^{-1}$ & $2.5\times10^{-1}$\\
        $^{32}$S & $2.3\times10^{-3}$ & $2.3\times10^{-3}$ & $2.4\times10^{-3}$ & $2.6\times10^{-3}$ & $1.4\times10^{-1}$ & $1.4\times10^{-1}$ & $1.4\times10^{-1}$ & $1.2\times10^{-1}$\\
\hline
    \end{tabular}
    \tablebib{(2)~\citet{gronow2021a}}
\end{table*}

\begin{table*}
    \caption{Final abundances for Model M11\_05\_001, M11\_05\_01, M11\_05\_1, and M11\_05\_3.}
    \label{tab:abund12}
    \centering
    \begin{tabular}{@{}lrrrrrrrr}
        \hline
        & \multicolumn{4}{c}{He detonation} & \multicolumn{4}{c}{core detonation} \\
        & M11\_05\_001 & M11\_05\_01 & M11\_05\_1$^{(2)}$ & M11\_05\_3 & M11\_05\_001 & M11\_05\_01 & M11\_05\_1$^{(2)}$ & M11\_05\_3 \\
        & [M$_\odot$] & [M$_\odot$] & [M$_\odot$] & [M$_\odot$] & [M$_\odot$] & [M$_\odot$] & [M$_\odot$] & [M$_\odot$]\\ \hline
        $^4$He & $1.1\times10^{-2}$ & $1.1\times10^{-2}$ & $1.0\times10^{-2}$ & $9.9\times10^{-3}$ & $9.2\times10^{-3}$ & $9.2\times10^{-3}$ & $8.4\times10^{-3}$ & $5.7\times10^{-3}$ \\
        $^{12}$C & $5.2\times10^{-6}$ & $5.3\times10^{-6}$ & $5.7\times10^{-6}$ & $5.7\times10^{-6}$ & $3.1\times10^{-6}$ & $3.1\times10^{-6}$ & $2.5\times10^{-6}$ & $1.2\times10^{-6}$\\
        $^{16}$O & $3.7\times10^{-3}$ & $3.7\times10^{-3}$ & $3.8\times10^{-3}$ & $4.0\times10^{-3}$ & $7.5\times10^{-4}$ & $7.5\times10^{-4}$ & $7.5\times10^{-4}$ & $7.6\times10^{-4}$\\
        $^{28}$Si & $5.5\times10^{-2}$ & $5.6\times10^{-2}$ & $5.6\times10^{-2}$ & $5.6\times10^{-2}$ & $4.5\times10^{-2}$ & $4.5\times10^{-2}$ & $4.6\times10^{-2}$ & $4.4\times10^{-2}$\\
        $^{32}$S & $2.5\times10^{-2}$ & $2.4\times10^{-2}$ & $2.4\times10^{-2}$ & $2.5\times10^{-2}$ & $3.8\times10^{-2}$ & $3.8\times10^{-2}$ & $3.7\times10^{-2}$ & $3.4\times10^{-2}$\\
\hline
    \end{tabular}
    \tablebib{(2)~\citet{gronow2021a}}
\end{table*}

\section{Metallicity-dependent nucleosynthetic yields}
\label{sec:results}

In the following, we discuss the effect different metallicities have on the
nucleosynthetic yields. For a comparison of the models at different masses
(e.g.,  all M10\_10 and M09\_05 models) as well as a discussion of the
differences in the detonation ignition mechanism of the various models we refer
to \citet{gronow2021a}. The nucleosynthetic yields are affected by the
modeling of the C detonation. The shock strength depends on its propagation
history. Because C detonations cannot be spatially resolved in SN simulations,
substantial modeling effort is necessary to settle this issue, which is beyond
the scope of this work. The relations of the nucleosynthetic yields of the
models found by \citet{gronow2021a} are the same for the models at the other
metallicities. Our analysis focuses on manganese as well as iron and nickel.
These elements originate in most part from SNe~Ia \citep{timmes1995a,
mcwilliam1997a,kobayashi2019b}, or are a main product of a SN~Ia explosion.

Tables~\ref{tab:abund01} to \ref{tab:abund12} list the nucleosynthetic yields
of selected isotopes for all models at $t=100$\,s after He detonation ignition.
They are given separately for the He detonation and core detonation. The
distinction is based on the initial He mass fraction of the cell a tracer is
associated with, thus cells with an initial He mass fraction larger than $0.01$
are considered to be part of the shell. The abundances of the models at solar
metallicity are taken from \citet{gronow2021a} with additional values given for
$^{54}$Fe, $^{55}$Fe, and $^{58}$Ni.  The yields of Model M2a are taken from
\citet{gronow2020a} and are extended by $^{52}$Fe, $^{54}$Fe, $^{55}$Fe,
$^{55}$Mn, $^{55}$Co, and $^{58}$Ni.

Detailed nucleosynthetic yields for the models at $0.01$, $0.1$, and
$3$\,$Z_\odot$ are given in the appendix in Tables \ref{app:stab0803_1} to
\ref{app:rad1105_1}. The yields of the solar metallicity models are included in
\citet{gronow2021a}. Tables \ref{app:stab0803_1} to \ref{app:stab1105_1} list
the abundances of stable nuclides, radioactive nuclides with lifetime longer
than $2\,\mathrm{Gyr}$ at time $t=100$\,s, and radioactive nuclides with
shorter lifetime decayed to stability. The nucleosynthetic yields of some
radioactive nuclides with lifetime shorter than $2\,\mathrm{Gyr}$ at $t=100$\,s
are given in Tables \ref{app:rad0803_1} to \ref{app:rad1105_1}. Each table
lists the abundances for one hydrodynamic explosion model at different
metallicities. The tables are only available in electronic form at the CDS.

\subsection{Nucleosynthesis in SNe~Ia}
The nucleosynthesis in SN~Ia explosions can be described as that of explosive
silicon (Si) burning with carbon and oxygen serving as fuel.
\citet{woosley1973a} distinguish three different burning regimes: incomplete Si
burning, $\alpha$-rich freeze-out and normal freeze-out from NSE.  Details of
the burning regimes and their leading reactions are also discussed by
\citet{lach2020a}. Figure 1 in \citet{lach2020a} illustrates the burning
regimes in the $T_\mathrm{peak} - \rho_\mathrm{peak}$\,-\,plane following
\citet{woosley1973a} for explosive Si burning. The burning regimes are also
marked in our Figures~\ref{fig:t-rho}, \ref{fig:he-t-rho}, and \ref{fig:t-rho2}
by shaded boundaries.

In double detonations of sub-M$_{\text{Ch}}$ CO WDs, explosive He burning is
taking place as well (\citealt{khokhlov1984a}; \mbox{\citealt{khokhlov1985a}}).
It is, however, clear by comparing Figure 1 of \citet{khokhlov1984a} \citep[see
also][]{khokhlov1985a} to our figures that NSE is not reached in the He
detonation of our models.

The generally low central density in sub-M$_{\text{Ch}}$ CO WDs is an important
parameter for the nucleosynthesis as it implies that the regime of normal
freeze-out from NSE is not reached. Instead, IGEs are produced in $\alpha$-rich
freeze-out and incomplete Si burning in the core detonation, and in the burning
of the He detonation.

\subsection{Low and intermediate mass elements}
The models show a varying impact of the metallicity on the nucleosynthetic
yields produced in the He and core detonations. This is the case because its
impact is different at higher densities present in the core. As described in
Section~\ref{sec:introduction}, the metallicity causes a neutron-excess in the
core while the isotopes of the metallicity implementation ($^{14}$N and
$^{22}$Ne among others) serve as seed nuclei for the reactions in the He
detonation and in the low density regime of the core \citep[see][for a
discussion of the effect on the detonation speed]{shen2014b}.  The latter
effect is important for the production of IMEs as they are mostly produced in
the incomplete Si burning regime at maximum densities lower than about
$2.5\times10^7$\,$\text{g cm}^{-3}$.  In the He detonation no or only little
influence of the metallicity is observed on the abundances of elements lighter
or equal to $^{44}$Ti. This is also the case for $^{12}$C, $^{16}$O, and
$^{28}$Si in the ejecta of the core detonation. The behavior is expected for
the fuel of the detonation (C and O) because the detonation propagation is not
affected by the metallicity as stated above. Also, IMEs such as $^{28}$Si are
not produced in NSE.  However, it is obvious from the yields produced in the
core detonation that an increase in metallicity decreases the $^{4}$He
abundance, which is a product in the $\alpha$-rich freeze-out. With increasing
neutronization the reaction
$^{4}$He($\alpha$n,$\gamma$)$^{9}$Be($\alpha$,n)$^{12}$C commences and even
becomes dominant in comparison to the triple-$\alpha$ reaction
\citep{howard1993b,hix1999a} supporting the burning of $^{4}$He.

\subsection{Manganese}
\label{sec:manganese}
$^{55}$Mn is the only stable isotope of manganese. It is produced directly in
incomplete Si burning and by the decay of $^{55}$Co via $^{55}$Fe. Since the
abundance of $^{55}$Mn usually is orders of magnitudes below that of $^{55}$Co
immediately after the explosive nuclear burning and prior to $^{55}$Co decay,
the production of Mn mainly depends on the $^{55}$Co production
\citep{truran1967a}. \citet{lach2020a} describe in more detail how the Mn
production is influenced by different parameters in the double detonation
scenario: the shell-core mass ratio, the density of the He shell and the
metallicity. The metallicity-effect is discussed in the following while
\citet{lach2020a} give a detailed account of the other effects.

In the He detonation, the impact of the metallicity increase on the $^{55}$Co
production is relatively weak. In contrast to that, the direct production of
$^{55}$Mn increases by one order of magnitude for each increase in metallicity
(from $0.01\,Z_\odot$ to $0.1\,Z_\odot$ to $1\,Z_\odot$ to $3\,Z_\odot$) .
Nevertheless, it only reaches values close to those of $^{55}$Co for Models
M09\_03\_3 and M08\_03\_3.  $^{55}$Fe is more abundant than $^{55}$Mn, and its
dependence on metallicity is similarly weak as that of $^{55}$Co as both are
less neutron-rich than $^{55}$Mn. The changes in the nucleosynthetic yields are
a result of the presence of $^{14}$N and, in addition, $^{22}$Ne, which was
mixed into the shell during the relaxation step of the hydrodynamic simulation.

The most important effect the metallicity has is the resulting neutron
excess and, therefore, lower $Y_e$ (due to $^{22}$Ne). This $Y_e$ is
approximately conserved during the duration of the explosive burning, which in
the case described here happens at time scales that are too short for
$\beta$-decays to change the electron fraction while densities are too low for
electron capture reactions to drive $Y_e$ lower. Effectively, the lower $Y_e$
reduces the free proton abundance during freeze-out, which leads to a net
decrease of the destruction of $^{55}$Co via $^{55}$Co(p,$\gamma$)$^{56}$Ni
which is the key reaction governing the final yield of $^{55}$Mn \citep[see
Table 13 of][]{bravo2012a}.

Figure~\ref{fig:t-rho} shows the tracer particle distribution of Models M10\_03
at four different metallicities (increasing from left to right) with
color-coded mass fraction of $^{55}$Mn (top) and $^{55}$Co (bottom). The trends
in the nucleosynthetic yields of $^{55}$Mn and $^{55}$Co which are produced in
the He detonation are visible. The tracer particles of the He detonation are
shown in Figure~\ref{fig:he-t-rho} for comparison. Furthermore,
Figure~\ref{fig:t-rho} illustrates that both isotopes, $^{55}$Mn and $^{55}$Co,
are produced in incomplete Si burning. As discussed by \citet{seitenzahl2013b},
$^{55}$Co is destroyed in $\alpha$-rich freeze-out from NSE by
$^{55}$Co($p,\gamma$)$^{56}$Ni. The production of $^{55}$Mn is, however,
several orders of magnitude lower than that of $^{55}$Co.

The amount of $^{55}$Co produced in the core detonation doubles from lowest to
highest assumed metallicity, while the production of $^{55}$Mn shows a steeper
increase with metallicity. The increase in $^{55}$Co is visible in the bottom
panel of Figure~\ref{fig:t-rho}: more $^{55}$Co is produced in the lower
temperature and density regime.  The presence of $^{22}$Ne has the same effect
on the nucleosynthesis of the core detonation as on the He detonation
($^{22}$Ne was mixed into the shell during the relaxation) since the isotopes
are produced in the same temperature and density region in both cases.
Generally, in both the He and core detonations, it is visible that the presence
of $^{22}$Ne, and with that the metallicity of the progenitor star, has a
strong impact on the production of Mn as suggested by \citet{seitenzahl2013b}.

Manganese is an element that is mostly produced in SNe~Ia with core-collapse
(CC) SNe contributing to the total production as well. It is not sufficiently
known yet what the origin of the solar [Mn/Fe] is. Different explosion
scenarios of SNe~Ia are usually considered to explain the rise in [Mn/Fe] at
[Fe/H] $\geq -1$ to solar values
\citep{matteucci1986a,cescutti2017d,eitner2020a,kobayashi2019b} seen in
observations \mbox{\citep{gratton1988a,gratton1991a}}.  \citet{seitenzahl2013b}
state that a source with a super-solar ratio of [Mn/Fe] is needed to explain
this trend. Their comparison of the Mn production from different progenitor
models leads them to conclude that at least 50\,\% of SNe~Ia originate from
near-M$_\mathrm{Ch}$ WD explosions. They argue that this is the case because
normal freeze-out is necessary to reach a high enough production of $^{55}$Co
so that [Mn/Fe] becomes super-solar. The production in incomplete Si burning is
not sufficient for this. However, normal freeze-out can only be achieved in
near-M$_{\text{Ch}}$ WDs which have high enough central densities.

\citet{seitenzahl2013b} examine WD mergers in their study to represent
explosions of sub-M$_{\text{Ch}}$ WDs and investigate a metallicity-dependence
for the near-M$_{\text{Ch}}$ WD explosion models. They argue that gravitational
settling is needed in order for $^{22}$Ne to be sufficiently abundant in the
high density regions to alter the direct $^{55}$Mn production. Since the effect
is small for near-M$_{\text{Ch}}$ WDs, it is neglected for sub-M$_{\text{Ch}}$
WD. In our models, we presume a homogeneous distribution of $^{22}$Ne in the
core under the assumptions that it was produced homogeneously during the
evolution of the zero-age main sequence progenitor star and that gravitational
settling has a minor effect. This leads to the presence of $^{22}$Ne affecting
the Mn production in the high density regime. Our study shows that [Mn/Fe] of
all models significantly increases with higher metallicity. This also applies
to [Mn/Fe] originating from the core detonation.  Accordingly, all models have
a super-solar [Mn/Fe] value at $3\,Z_\odot$ (see bottom panel in
Figure~\ref{fig:metal}). The contribution of progenitors with super-solar
metallicity to the solar Mn over Fe ratio at [Fe/H]$=0$ is not well known. They
do not contribute to the ratio at that point in one-zone galactic evolution
models.  \citet{lach2020a} show that [Mn/Fe] is significantly super-solar in
the nucleosynthetic yields of the shell detonation increasing the total Mn over
Fe ratio. Thus showing that the He detonation plays an important role for the
total [Mn/Fe]. A comparison of the $^{55}$Mn and $^{55}$Co yields originating
from the shell and core detonation of Models M08\_10, M08\_05, and M08\_03
(Tables~\ref{tab:abund09}, \ref{tab:abund10}, and \ref{tab:abund11}) confirms
that the contribution of the shell detonation becomes more relevant at higher
shell masses. The relation is the same among other models with equal core mass.

We point out that \citet{seitenzahl2013b} use CC SN yields provided by
\citet{woosley1995a} in their GCE model. These have a strong influence on the
predicted evolution of [Mn/Fe] over [Fe/H]. They are similar to those by
\citet[N13]{nomoto2013a}. The chemical evolution calculations presented in
\citet{seitenzahl2013b} would potentially lead to a less stiff requirement for
M$_{\text{Ch}}$ WD explosions if CC SN yields by \citet[LC18]{limongi2018a}
were used.

The effect of core-shell mixing on [Mn/Fe] is investigated by a comparison of
Models M08\_10 by \citet{gronow2021a}. While one of the models has the core and
shell compositions obtained after the relaxation step, the mixing is reset in
the other model (see \citealp{gronow2021a} for a detailed description of both
models). The models have solar metallicity.  [Mn/Fe] is about $0.2$ in both
cases with a difference of a few percent indicating that the mixing caused by
the relaxation only has a small effect on the value. Variations in the
$^{55}$Mn production are minor as well. Some admixture of C into the shell in
the initial transition region rather already influences the yields as described
in \citet{yoon2004b} and \citet{gronow2020a}.

\subsection{Iron and nickel}
\label{sec:ni}
$^{56}$Ni is the main product of a SN~Ia explosion. Thus, the iron abundances
is high as well after the decay of $^{56}$Ni. Among those elements the isotopes
$^{54}$Fe and $^{58}$Ni are interesting as they are the next stable isotopes to
the $\alpha$-chain elements $^{52}$Fe and $^{56}$Ni. Further, $^{57}$Ni can be
measured from observations \citep{graur2016a} via its decay to $^{57}$Co
allowing the determination of $^{57}$Ni/$^{56}$Ni.

An increase of the abundances originating from the He detonation with
metallicity, however not as strong as for $^{55}$Mn and $^{55}$Co, can also be
observed for $^{54}$Fe and $^{58}$Ni. In general, a higher metallicity leads to
an increased production of stable IGE nuclides. This can also be observed in
the core detonation (see Tables~\ref{tab:abund01} to \ref{tab:abund12}). The
abundances of $^{54}$Fe and $^{58}$Ni at $3$\,$Z_\odot$ increase to four times
the value found at $0.01$\,$Z_\odot$.  The change in the nucleosynthetic yields
is due to the presence of additional neutrons which stem from $^{22}$Ne via
$^{22}$Ne($\alpha$,n)$^{25}$Mg in the $\alpha$-rich freeze-out regime
\citep{shigeyama1992a}. The reaction produces the neutron-rich $^{25}$Mg.  In
subsequent ($\alpha$,n) and (n,$\gamma$) reactions other neutron-rich isotopes
are formed up to IGEs, such as $^{54}$Fe and $^{58}$Ni which have two extra
neutrons compared to the $\alpha$-chain elements $^{52}$Fe and $^{56}$Ni.
Additionally, the neutron, that is freed in the reaction of $^{22}$Ne to
$^{25}$Mg, allows the production of C isotopes in reactions with $^{20}$Ne. In
this context free p are captured in exchange for $\alpha$-particles
\citep{chamulak2007a}. Contrary to this effect, the abundances of Fe and Ni
which are produced in the He detonation are changed as $^{22}$Ne and other
isotopes alter the leading reactions. These isotopes compose the metallicity in
the simulations and serve as seed nuclei for the reactions. As such
$^{14}$N($\alpha$,p)$^{17}$O can lead to a speed up of the burning.
The material undergoing explosive burning quickly reaches nuclear
statistical quasi-equilibrium (QSE), which comprises two equilibrium clusters,
around the Si and Fe groups \citep[see also][]{lach2020a}. The clusters are
separated by a bottleneck at A $\approx 45$ (Ca/Ti/Sc). This boundary of the QSE
clusters depends on $Y_e$ and the main nuclear reactions that connect the Si
QSE cluster with the Fe-group cluster shift as a function of the neutron
excess. At $Y_e\approx0.5$ the bottleneck is by-passed with reactions such as
the most important bridging flow through $^{45}$Sc(p,$\gamma$)$^{46}$Ti, but
also $^{42}$Ca($\alpha$,$\gamma$)$^{46}$Ti and $^{45}$Ti(n,$\gamma$)$^{46}$Ti
\citep{woosley1973a} and $^{44}$Ti($\alpha$,p)$^{47}$V \citep{bodansky1968a}.
Instead, ($\alpha$,n) and ($\alpha$,$\gamma$) reactions on argon isotopes and
(p,n) reactions on potassium isotopes, such as $^{40}$Ar($\alpha$,n)$^{43}$Ca,
$^{38}$Ar($\alpha$,$\gamma$)$^{42}$Ca or $^{42}$K(p,n)$^{42}$Ca, dominate the
fluxes upwards out of the Si-group at $Y_e\approx0.46$.
$^{43}$Ca(n,$\gamma$)$^{44}$Ca followed by a series of (n,$\gamma$) reactions
along the Ca-isotopic chain up to $^{48}$Ca then dominate the flux into the
Fe-group QSE cluster \citep{hix1996a}. A consequence of this shift is the
increasd production of neutron rich Fe-group isotopes, like $^{54}$Fe and
$^{58}$Ni, in incomplete Si burning.

\begin{figure*}
    \centering
    \includegraphics[width=0.98\textwidth]{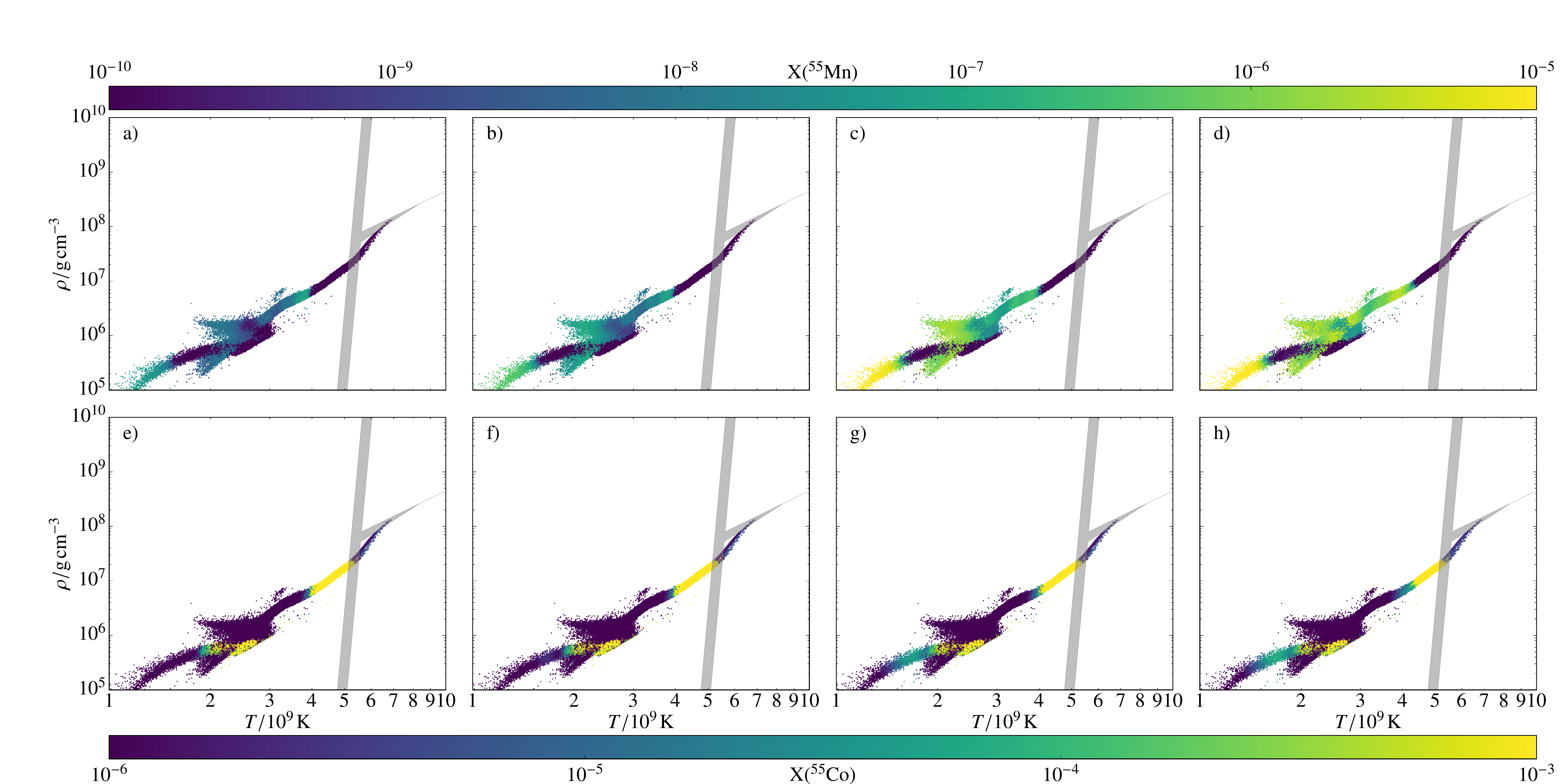}
    \caption{Tracer particle distribution in the $T_\mathrm{peak} -
        \rho_\mathrm{peak}$\,-\,plane for Models M10\_03 at $0.01$, $0.1$, $1$, and
        $3$\,$Z_\odot$, mass fractions of $^{55}$Mn (top, a) to d)\,) and $^{55}$Co
    (bottom, e) to h)\,) at $t=100\,$s are color coded. The shaded areas
split the domain into normal freeze-out from NSE, $\alpha$-rich freeze-out and
incomplete Si-burning (clockwise starting in the top right) following
\citet{woosley1973a} \citep[see also][]{lach2020a}.}
\label{fig:t-rho}
\end{figure*}

\begin{figure}
    \centering
    \includegraphics[width=0.46\textwidth]{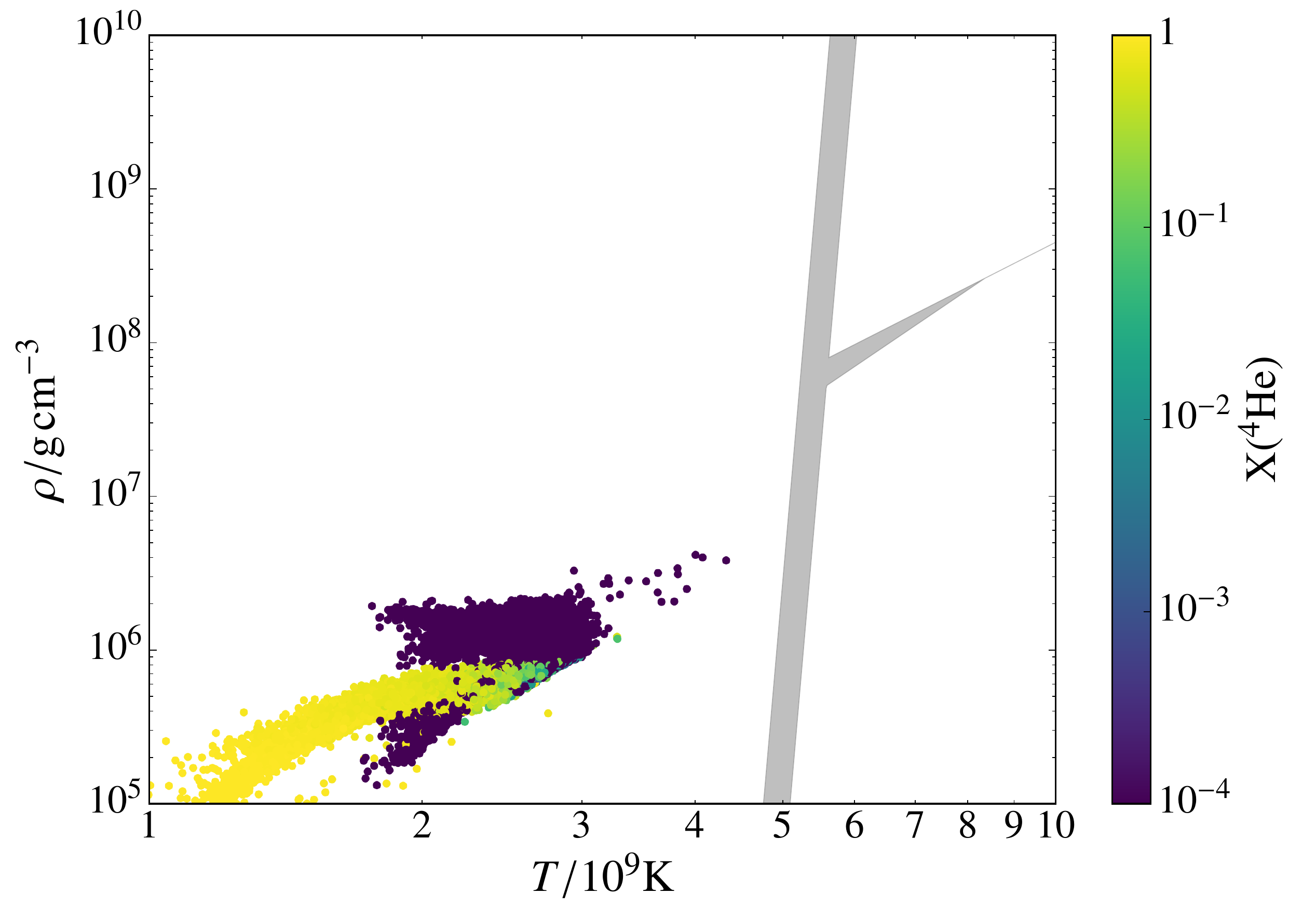}
    \caption{Tracer
    particle distribution in the $T_\mathrm{peak} - \rho_\mathrm{peak}$\,-\,plane
of the He detonation for Model M10\_03\_001 at $t=100\,$s. The $^{4}$He mass fraction
is color coded. The shaded areas are the same as in Figure~\ref{fig:t-rho}.}
\label{fig:he-t-rho}
\end{figure}

\begin{figure}
    \centering
    \includegraphics[width=0.49\textwidth]{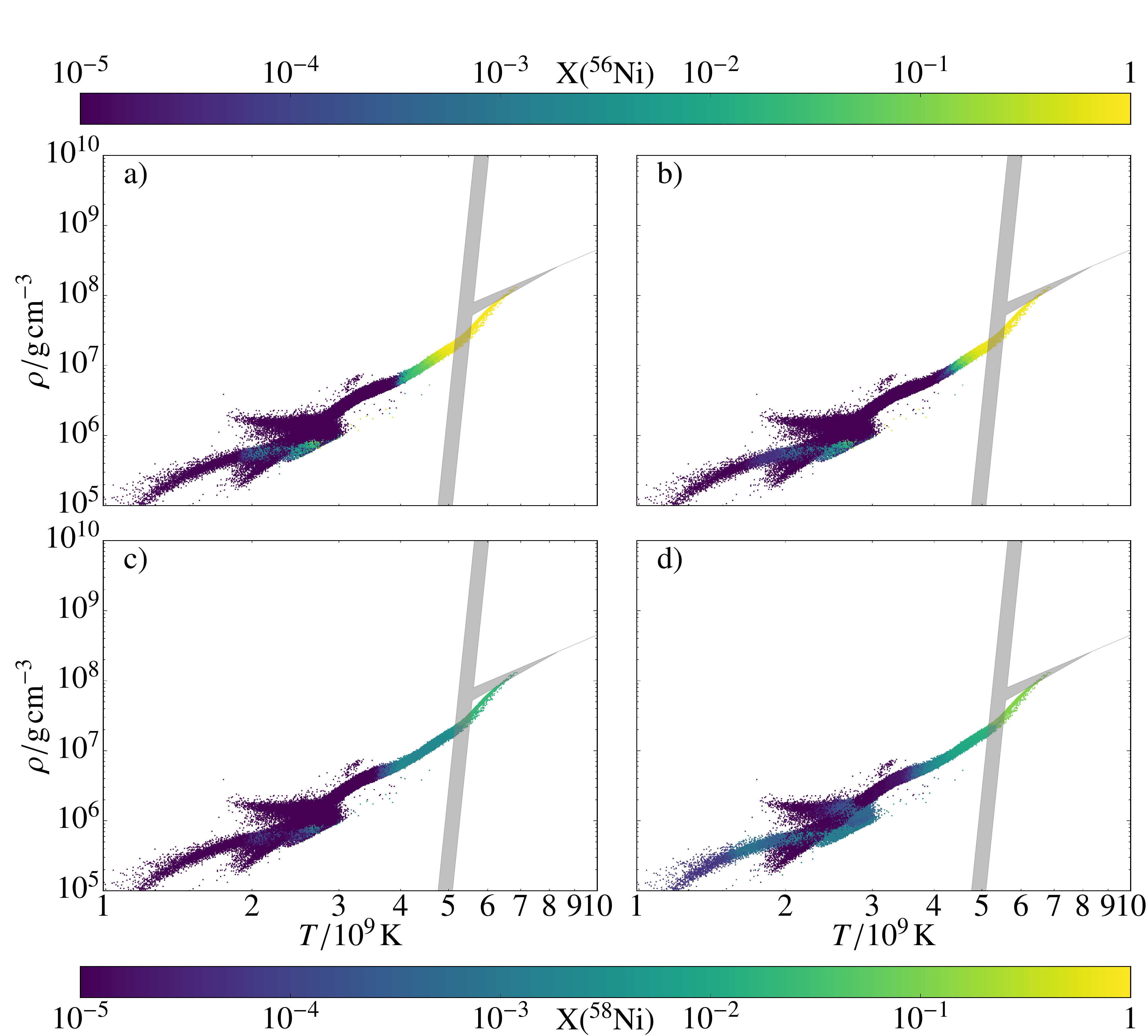}
    \caption{Tracer particle distribution in the $T_\mathrm{peak} -
    \rho_\mathrm{peak}$\,-\,plane for Models M10\_03\_001 (left) and M10\_03\_3, mass
fractions of $^{56}$Ni (top) and $^{58}$Ni (bottom) at $t=100\,$s are color
coded with shaded areas as in Figure~\ref{fig:t-rho}.}
\label{fig:t-rho2}
\end{figure}

The production of $^{56}$Ni and $^{58}$Ni is illustrated in
Figure~\ref{fig:t-rho2} in the same way as in Figure~\ref{fig:t-rho} for Model
M10\_03\_001 (left) and Model M10\_03\_3 (right). A comparison shows that the
production of $^{56}$Ni at lower densities in the incomplete Si burning regime
decreases with higher metallicity, though the differences are only weakly
pronounced. The same applies to $^{56}$Ni produced in the He detonation.
Changes in the $^{58}$Ni abundances are more visible in
Figure~\ref{fig:t-rho2}. The differences in the $^{58}$Ni yields are most
prominent in the density regime below $3\times10^6$\,$\text{g cm}^{-3}$ while a
small increase is visible in the $\alpha$-rich freeze-out regime as well.

\begin{figure}
    \centering
    \includegraphics[width=0.49\textwidth]{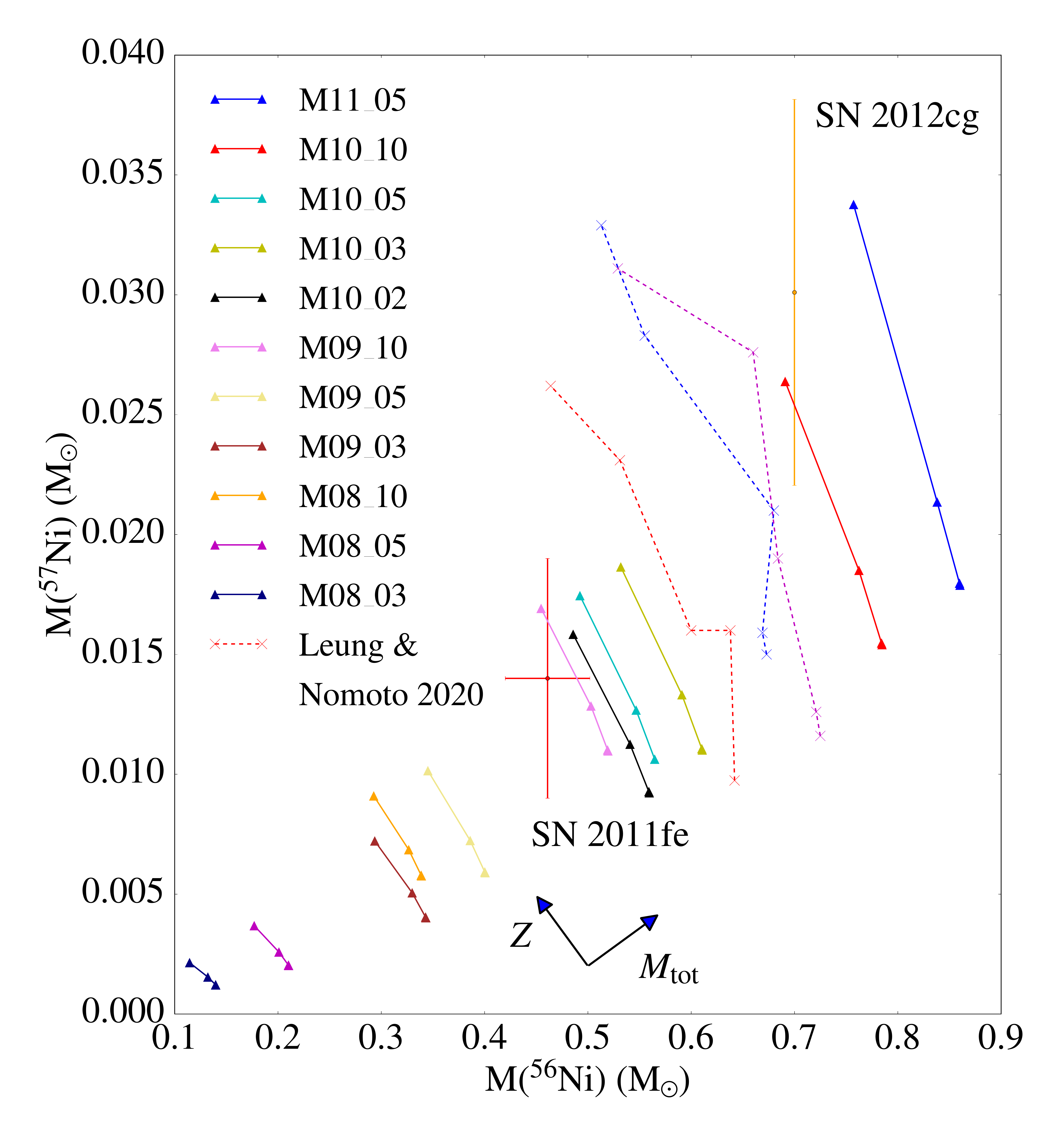}
    \caption{Mass of $^{57}$Ni over $^{56}$Ni mass for all models (solid lines). Models of
    \citet{leung2020a} are added for comparison (dashed lines). The data for
SN\,2011fe is taken from \citet{dimitriadis2017a} (case 1) and for SN\,2012cg
from \citet{graur2016a}.}
\label{fig:ni56-57}
\end{figure}

\citet{timmes2003a} point out that the metallicity of the WD influences the
electron fraction $Y_e$ (an increase in metallicity decreases $Y_e$). A
decrease in $Y_e$ in turn increases the nucleosynthetic yields of $^{54-58}$Fe
and $^{57-60}$Ni. This is in agreement with \citet{curtis2018a} who state that
a lower $Y_e$ supports the production of $^{57}$Ni over $^{56}$Ni, thus
changing the $^{57}$Ni/$^{56}$Ni ratio. This is confirmed in our simulations
(see Figure~\ref{fig:ni56-57}). Shown is the total $^{57}$Ni mass over
$^{56}$Ni mass for all models. Models with the same mass configuration (e.g.,
all M08\_03 models) have the same color and are connected by solid lines. The
metallicity increases to the top left and the total mass of the models
increases to the top right. Two models (Models M08\_10 and M10\_03) have higher
$^{56}$Ni values than the model with the next highest total mass which is due
to the different configurations of the models. For example, the shell mass of
Model M08\_10, along with the densities present in the shell, allows a
production of $^{56}$Ni in the He detonation which is of the order of
$10^{-2}\,M_\odot$.  This is significantly higher than for Model M09\_03 and
more than balances the small difference of the $^{56}$Ni production in the core
detonation.  It is visible that the $^{56}$Ni decreases and the $^{57}$Ni
increases with increasing metallicity.  The yields at $0.01\,Z_\odot$ and
$0.1\,Z_\odot$ show only small differences which are marginally visible for
Models M11\_05\_001 and M11\_05\_01, but overlap for all other models. A
comparison to SN\,2011fe \citep[case 1 of][]{dimitriadis2017a} shows that it
lies in the range covered by our models.  It suggests that Model M09\_10\_3 can
be a good fit and that a WD with this total mass (and possibly mass
configuration of core and shell) can be its progenitor.  Furthermore,
SN\,2012cg \citep{graur2016a} could be explained by a progenitor similar to
Model M10\_10.

\subsection{Nucleosynthetic yields relative to solar values}
\begin{figure*}
    \centering
    \hspace*{-1.0cm}
    \vspace*{-0.58cm}
    \includegraphics[width=1.1\textwidth]{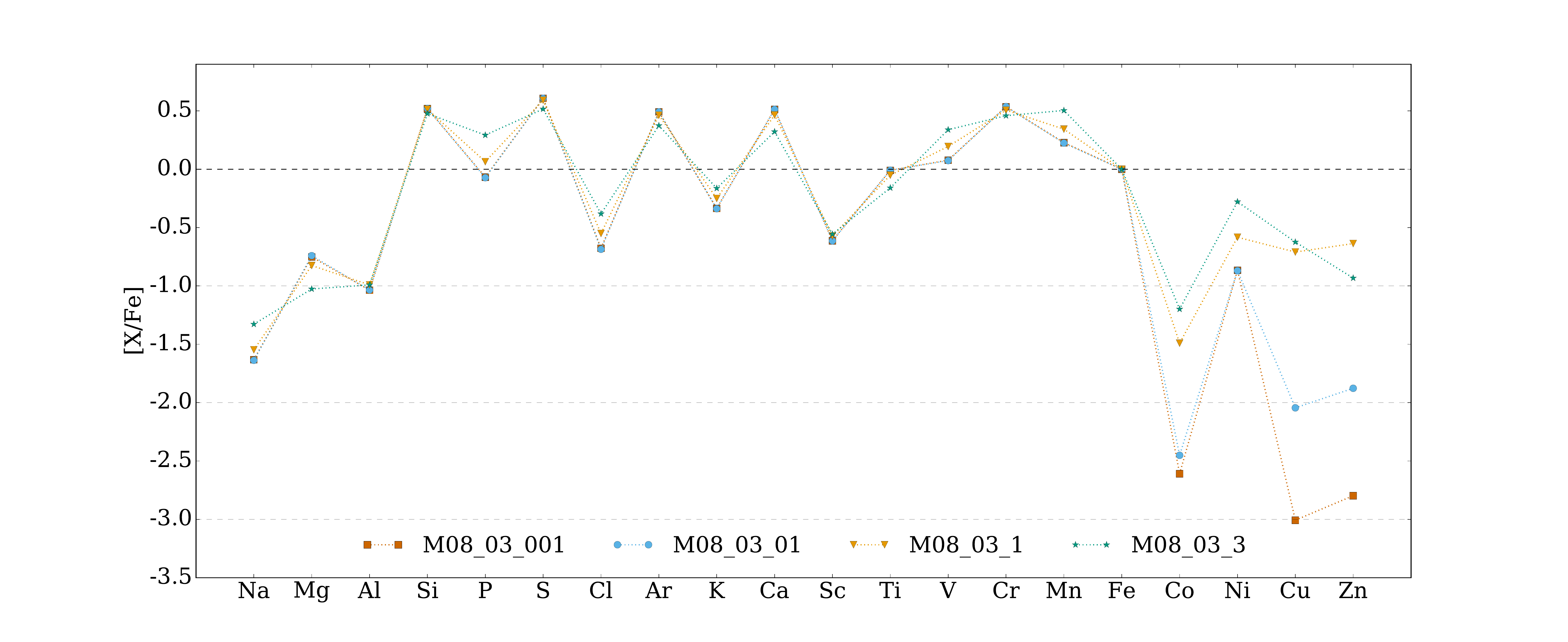}
  \vspace*{-0.5cm}
  \hspace*{-1.0cm}
      \includegraphics[width=1.1\textwidth]{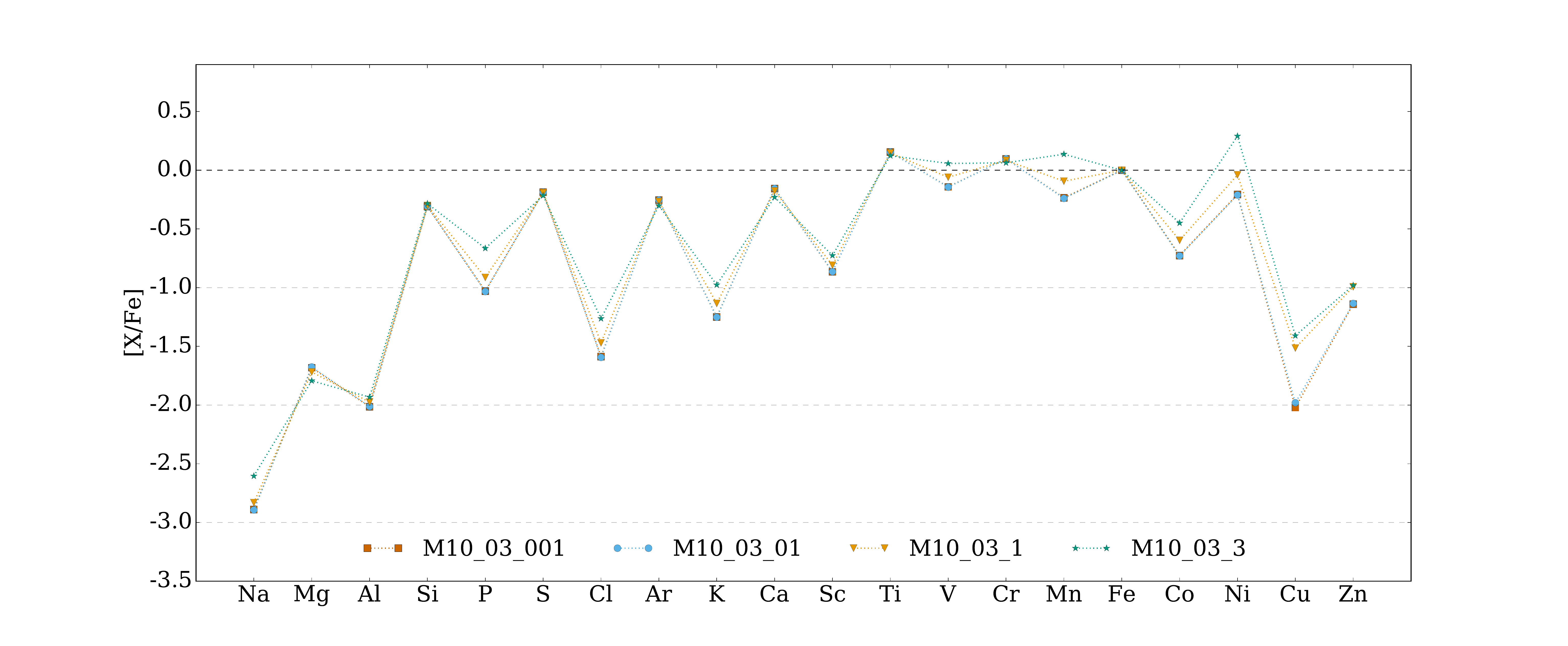}
  \caption{Elemental ratios relative to Fe compared to solar ratios for Models M08\_03
  (top) and M10\_03 (bottom) at four different metallicities.}
  \label{fig:model}
\end{figure*}

\begin{figure*}
  \hspace*{-1cm} \includegraphics[width=1.1\textwidth]{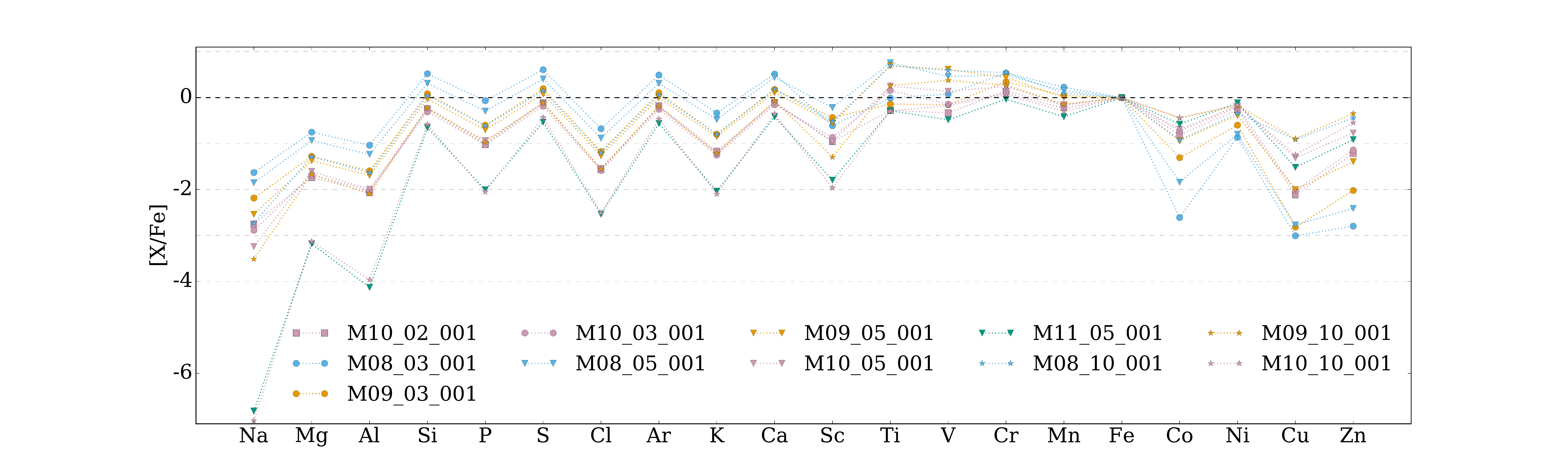}
  \hspace*{-1cm} \includegraphics[width=1.1\textwidth]{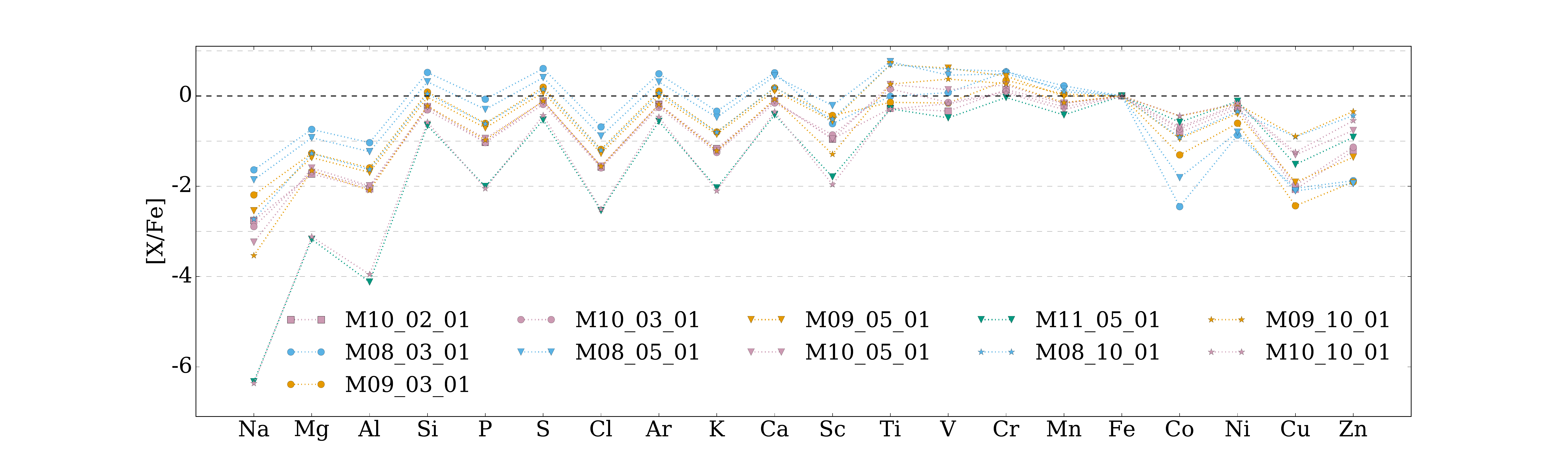}
  \hspace*{-1cm} \includegraphics[width=1.1\textwidth]{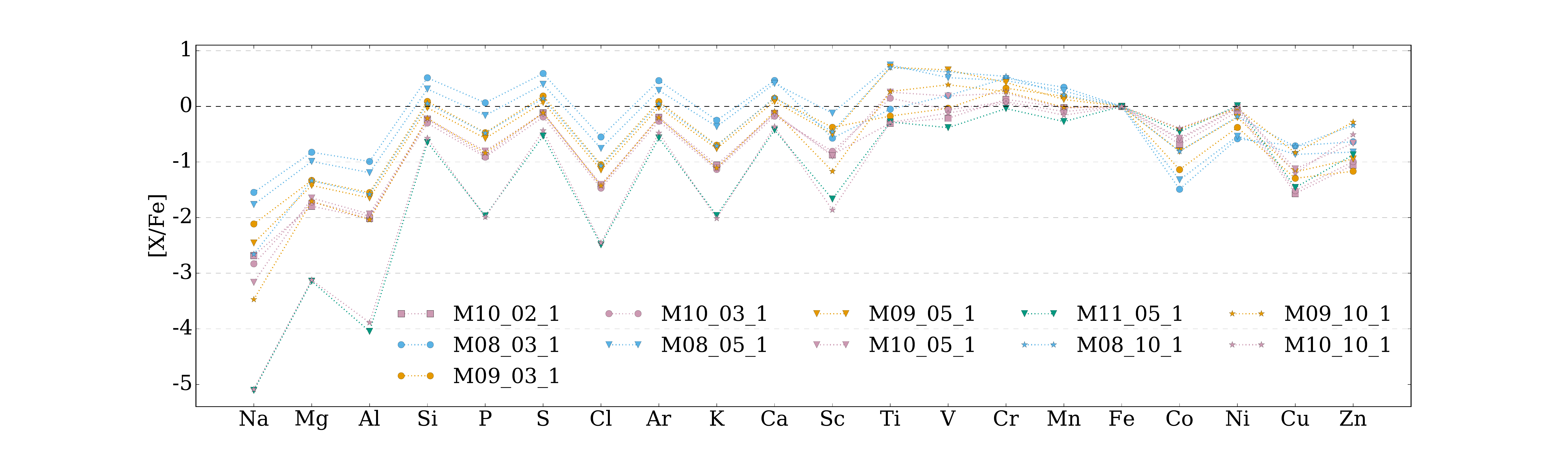}
  \hspace*{-1cm} \includegraphics[width=1.1\textwidth]{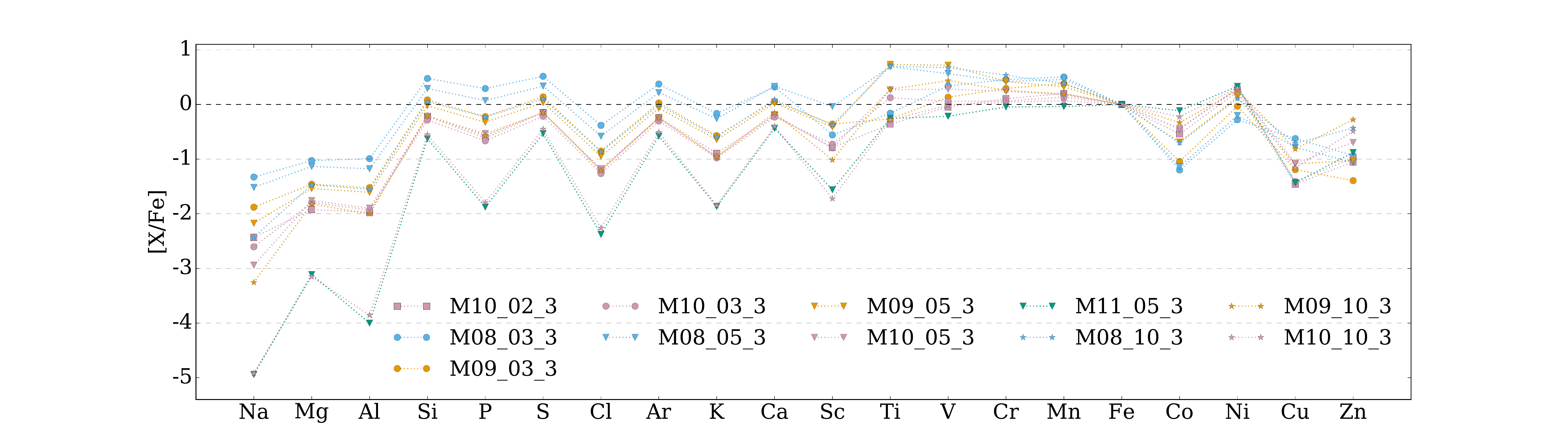}
  \caption{Elemental ratios relative to Fe compared to solar ratios for all
  models sorted by metallicity, from top to bottom: $0.01$\,$Z_\odot$,
  $0.1$\,$Z_\odot$, $1$\,$Z_\odot$, and $3$\,$Z_\odot$.}
  \label{fig:metal}
\end{figure*}

The variation of the nucleosynthetic abundances is apparent in
Figures~\ref{fig:model} and \ref{fig:metal}. Both figures give the elemental
ratio relative to iron with respect to solar ratios \citep{asplund2009a}. For
this, radioactive nuclides are decayed to $2\,\mathrm{Gyr}$. Figure
\ref{fig:model} shows the elemental ratios for Models M08\_03 (top) and M10\_03
(bottom) at four different metallicities, while Figure \ref{fig:metal}
illustrates the elemental ratios of all models sorted by metallicity. The
models with the lowest metallicity of $0.01$\,$Z_\odot$ at the top and at
$3$\,$Z_\odot$ at the bottom.

The influence of the metallicity on the elemental ratios is visible in
Figure~\ref{fig:model}. The two panels show similar trends for both models.
Among those is the previously discussed increase in Mn with increasing
metallicity. It can further be seen that the Ti production decreases (a little)
with increasing metallicity along with Cr. Generally, this decrease indicates a
better match to observables as Ti and Cr often cause discrepancies in early
observations \citep{hoeflich1996b,kromer2010a}.  However, the decreasing values
here are caused by changes in the nucleosynthetic yields coming from the core
detonation and not the He detonation. An improved reproduction of observations
is therefore not expected. One further trend is the odd-even effect in the
production of intermediate mass elements (IMEs) \citep[see][and references
therein]{peterson1981a}. Elements with an even atomic number are mostly
produced in the $\alpha$-chain. The pairing effect results in a stable
structure due to the high nuclear binding energy. Contrary to that, the nuclear
binding energy of elements with an odd atomic number is low. The production of
these elements further depends on the neutron excess of the white dwarf
\citep{wheeler1989a} which is visible in Figures~\ref{fig:model} and
\ref{fig:metal}. This is in most part due to the fact that stabilities shift
toward neutron-rich isotopes for heavier elements.

A drop-off in the elemental ratios is noticeable for copper and zinc while
cobalt is produced at sub-solar values as well. However, the effect decreases
with increasing metallicity.  Models with a more massive He shell further tend
to reach higher values of Co, Cu, and Zn than the respective comparison models
with lower shell masses at the same metallicity (see Figure~\ref{fig:metal}).
This confirms that Cu and Zn are mostly produced in the He detonation.  The
sub-solar production of the three elements Co, Cu, and Zn is typical for pure
detonation models as pointed out by \citet{lach2020a}. Generally some of these
features are present in all models as shown in Figure~\ref{fig:metal} and the
relations between the models mostly stay the same with changing metallicity. A
comparison of models with the same core mass at the various metallicities
further allows to detect the influence of the He detonation on the elemental
ratios which is already discussed in \citet{lach2020a} and confirmed here.

The super-solar production of Ti, V, and Cr is caused by the He detonation.
However, the values are almost solar for Models M08\_03, M09\_03, and M10\_03.
Especially Model M10\_03 is a good fit at all metallicities. The other models
match Ti and V better than Cr and also show larger deviations toward higher
metallicity.  The Mn production is almost solar with an increasing trend at
higher metallicity. At all metallicities it is visible that the Mn to Fe ratio
is lower with increasing core mass as a high [Mn/Fe] stems from the Mn
production in the shell. The relative contribution of Mn coming from the He
detonation decreases with increasing core mass. Additionally, more Fe is
produced at the higher densities in WDs with higher-mass cores because burning
extends to higher-mass elements in these cases. Therefore, at solar metallicity
only few models with sufficient mass to produce normal SNe~Ia reach solar
[Mn/Fe] (Models M09\_05\_1, M09\_10\_1, and M10\_05\_1). However, all models
have super-solar values at a metallicity of $3$\,$Z_\odot$.

\subsection{Impact on observables}
We have carried out radiative transfer simulations for a representative sub-set
of models (Models M10\_03) to investigate the influence of metallicity on the
predicted observables. These will be presented in detail in a follow-up paper.
However, we discuss the results qualitatively here.

Low metallicities ($\leq Z_\odot$)  do not have a significant impact on the
predicted photospheric-phase light curves and spectra.  At early times, double
detonation simulations tend to show colors too red compared to observations of
normal SNe~Ia, predominantly due to the production of heavy elements in the
shell detonation \citep[Cr, Ti, and IGEs, e.g.,][]{kromer2010a,
gronow2020a}.  As discussed above, increasing the metallicity does not
significantly impact the production of these elements in the shell detonation.
Therefore, it is not unexpected that the metallicity of the model does not
substantially alter the agreement with observations for low to moderate values.
At high metallicity ($3$\,$Z_\odot$), there is a noticeable impact on the
predicted observables, such that the models are fainter and redder.  This can
be attributed to the lower masses of $^{56}$Ni that are produced at this high
metallicity, leading to lower ejecta temperatures and redder colors.

Although the impact on the early observables is relatively modest, metallicity
can become more important when considering late times.  This is because the
metallicity affects the ratio of stable to radioactive isotopes produced in the
model.  In the nebular phase, the relatively rapid decay of $^{56}$Ni means
that $^{58}$Ni becomes the dominant Ni isotope. Therefore, the abundance of
this stable isotope can be directly probed in nebular spectra
\citep{ruiz-lapuente1992b}.  Increasing metallicity provides a mechanism by
which sub-M$_{\text{Ch}}$ models can produce increasing amounts of stable
Ni, although it remains unclear whether conditions in sub-M$_{\text{Ch}}$
models can yield [\ion{Ni}{II}] emission to the degree required by data
\citep{shingles2020a, wilk2020a}.

\section{Comparison to previous work}
\label{sec:discussion}
Other models involving a non-zero metallicity of the zero-age main sequence
progenitor were investiagted by \citet{shigeyama1992a, timmes2003a, sim2010a,
shen2018b}, and \citet{leung2020a}.  In the following we compare our models to
theirs where possible. We point out that only \citet{leung2020a} examine a
sub-M$_{\text{Ch}}$ WD with a He shell with all others studying detonations in
bar sub-M$_{\text{Ch}}$ CO WD.  A discussion of the models at solar metallicity
was already carried out in \citet{gronow2021a}.

\citet{lach2020a} discuss the nucleosynthesis results of Model M10\_05\_1
(Model M2a$_\odot$ in their paper) in connection with other explosion
scenarios, such as delayed detonations or pure deflagrations, also analyzing
their impact on GCE. A specific focus is given on Mn, Zn, and Cu. Similar to
other work \citep[e.g.,][]{seitenzahl2013b}, they confirm metallicity as an
important parameter for the production of Mn which we also find in the more
detailed metallicity study presented here. We further confirm that the Cu
production increases with metallicity though it is still underproduced compared
to solar values also in our highest metallicity models.

\citet{floers2019a} use super-solar values of Mn and Ni to differentiate
between M$_\mathrm{Ch}$ and sub-M$_\mathrm{Ch}$ WDs as progenitors in nebula
spectra. As pointed out by \citet{lach2020a} the distinction between both
models is not that straightforward. At high metallicity ($1$\,$Z_\odot$) our
models with a core mass of at least $1.0\,M_\odot$ reach at least solar values
for both elements.

\citet{sim2010a} explore detonations in a variety of sub-M$_{\text{Ch}}$ CO WDs
at different masses.  Only one of the models is calculated at about
$3$\,$Z_\odot$. This model has a total mass of $1.06\,M_\odot$ which is similar
to our Models M10\_05. However, their model does not have a He shell. They find
that more stable IGEs are produced compared to the same model at zero
metallicity.  Figure 3 of \cite{sim2010a} also shows that less $^{56}$Ni is
synthesized in the inner and less IMEs in the outer regions. These changes are
attributed to the presence of $^{22}$Ne which is used to represent metallicity
in the WD. While it is not possible to compare the respective detailed
nucleosynthetic yields, our models agree with the trends found by
\citet{sim2010a}.

A similar model (WD without He shell) is calculated by \citet{shen2018b}.
However, they carry out a parameter study involving different WD masses and
metallicities. Similar to us, they examine four different metallicities of the
main sequence progenitor star, namely $0$, $0.5$, $1$, and $2$\,$Z_\odot$. They
use $^{22}$Ne and $^{56}$Fe as proxies for the metallicity in the hydrodynamic
simulation.
Similar to us they find that the production of $^{56}$Ni decreases with
increasing metallicity while other, stable IGEs are produced. A
rough comparison of the \citet{shen2018b} models can be carried out to the
yields originating from the core detonation in our models with the least
massive He shells. Generally, the same behavior is found in all models. The
yields of some isotopes, such as $^{32}$S and $^{40}$Ca, are within a few
perent of each other at $0.01\,Z_\odot$ (compared to zero metallicity of
\citet{shen2018b}) and $Z_\odot$. \citet{shen2018b} compare their solar
metallicity models to the ones of \citet{blondin2017a} and argue that the
discrepancy in the $^{56}$Ni abundance is caused by the smaller nuclear
reaction network used by \citet{blondin2017a}. We point out that the
differences in the $^{56}$Ni abundance reach up to two-thirds of the values by
\citet{shen2018b} depending on the WD mass.  The models by \citet{shen2018b},
however, match the models by \citet{shigeyama1992a} relatively well at
$2$\,$Z_\odot$.

The models by \citet{shigeyama1992a} are calculated in 1D and involve
sub-M$_{\text{Ch}}$ CO WDs without a He shell. Similar to us they find an
increase in the production of neutron-rich IGEs due to the presence of
$^{22}$Ne which they use to approximate metallicity in the same way as done in
our study.  A more detailed comparison is not feasible due to the different
setups of the models.

\citet{timmes2003a} predict a relation between the amount of produced $^{56}$Ni
and the metallicity of the white dwarf. Their 1D models of M$_\mathrm{Ch}$ WDs
show a decrease of $^{56}$Ni by 25\% going from $0.3$\,$Z_\odot$ to
$3$\,$Z_\odot$.  \citet{shen2018b} find a similar trend when the metallicity is
changed from 0 to 2\,$Z_\odot$ but the amount of the reduction varies with WD
mass in their case. Models with a mass of $0.8\,M_\odot$ show a decrease of
50\% in $^{56}$Ni and models of a $1.0\,M_\odot$ WD only of 10\%.
\citet{bravo2019b} carry out a parameter study involving
sub-M$_{\text{Ch}}$ CO WDs with masses between $0.88\,M_\odot$ and
$1.15\,M_\odot$ at varying metallicities. Their $^{56}$Ni abundances show a
similar decrease by about 17\%. Our models agree with the trend found in these
previous studies. The decrease, however, is lower with values from 12\% to
21\%. While the difference of the $^{56}$Ni abundance at $0.01$\,$Z_\odot$ and
$3$\,$Z_\odot$ is of the same order for most models, our model with the
smallest total mass shows the highest value of 21\%.  The derived linear
relation of \citet{timmes2003a} assumes a fixed $^{56}$Ni mass of
$0.6\,M_\odot$ produced in a normal SN~Ia. However, this mass is in large part
influenced by the total mass of the white dwarf. Lower-mass stars produce
significantly less $^{56}$Ni as can be seen in Tables~\ref{tab:abund01} to
\ref{tab:abund12}.  The relative change of this mass with increasing
metallicity is therefore larger at lower total masses.

\cite{seitenzahl2013a} and \citet{leung2018a} carry out studies involving
different metallicities for near-M$_\mathrm{Ch}$ WDs. A comparison of the
models is therefore only possible analog to the comparison with the studies of
\citet{sim2010a} and \citet{shen2018b}.  Instead we focus on a study by
\citet{leung2020a} who investigate double detonations of sub-M$_{\text{Ch}}$
WDs. They look into similar core and shell masses as presented in our paper.
However, our models reach down to lower core and shell masses. In contrast to
our models, \citet{leung2020a} do not consider any mixing between core and
shell. This has an influence on the nucleosynthetic yields produced in the He
detonation as shown by \mbox{\citet{gronow2020a}}.  The metallicities they
include in their study reach from $0$\,$Z_\odot$ to $5$\,$Z_\odot$ which is
approximated by $^{22}$Ne similar to our hydrodynamic models. The metallicity
is treated more accurately in the postprocessing step of our models as $^{14}$N
is included as a proxy for the metallicity in the He shell of the hydrodynamic
simulation while \citet{leung2020a} do not include this in the
hydrodynamic and postprocessing calculations (Leung, priv. comm.). In
addition, the solar abundances of \citet{asplund2009a} are scaled to the
respective metallicity as described in Section~\ref{sec:models_methods}.
\citet{lach2020a} compare an approach that only includes $^{22}$Ne in
the postprocessing calculation as proxy for the metallicity, such as used by
\citet{leung2020a} and other authors before, with the one presented here as
well as in their study. The simulations by \citet{leung2020a} are only
carried out in 2D and cover one quarter of the star. Furthermore, their
hydrodynamic simulations only include seven isotopes in the nuclear reaction
network. This is not best suited to cover the energetics of the shell
detonation as pointed out by \citet{shen2018b} and \citet{townsley2019a}.  The
nuclear network used in the hydrodynamic models of \citet{gronow2021a}
comprises 55 isotopes to match the one of \citet{townsley2019a}. This allows to
cover the He and Si burning more accurately than with a small nuclear network.
Additionally, the level-set method used in \citet{leung2020a} is not optimal to
follow the detonation front at low densities present in the shell and in
low-mass WDs in general \citep{gronow2020a, gronow2021a}. The energy release
and nuclear burning are not calculated self-consistenly using this method
\citep{gronow2020a}. Instead the energy release in the burning is treated in a
parametric way which requires a calibration for the simulation
\citep{fink2010a}. The \textsc{Arepo} code instead enables a coupling
of the nuclear network to the hydrodynamics \citep{pakmor2013a}.  The groups H,
I, and J of \citet{leung2020a} focus on the effect of the metallicity on the
explosion characteristics and nucleosynthetic yields. Out of these models,
Group I is most similar to Model M10\_10 as it has a total mass of
$1.1\,M_\odot$ with a $0.1\,M_\odot$ shell.  \citet{leung2020a} find that the
production of $^{56}$Ni decreases with increasing metallicity. Similar to
\citet{timmes2003a} they find a decrease of about 20\%. They conclude that the
metallicity has only little to no impact on the explosion energy, final energy,
detonation channel, detonation position and time of core ignition. This
supports our computational approach in which we do not calculate the
hydrodynamic simulation for each of our 44 models, but use the hydrodynamic
simulation of eleven and follow-up on those with detailed nucleosynthesis
calculations at different metallicities in a postprocessing step. A detailed
analysis of the models of Group I in \citet{leung2020a} shows an increase in
the production of stable isotopes (both IMEs and IGEs) while the final yields
of the $\alpha$-chain elements are not affected by the metallicity. The
difference of the neutron-rich isotopes can be as high as four orders of
magnitude as stated by \citet{leung2020a}. We find a similar increase, most
significantly in the yields of $^{55}$Mn as discussed in Section
~\ref{sec:manganese}. Tables 6 and 7 of \citet{leung2020a} list nucleosynthetic
yields of stable and some radioactive isotopes of the benchmark model in Group
I. Here seven different metallicities are included out of which three are part
of our study. Comparable yields of Model M10\_10 are given in the appendix and
the appendix of \citet{gronow2021a}. A comparison of the abundances shows that
the $^{56}$Ni production is higher in our models though the decreasing trend
with metallicity is similar. Other isotopes confirm that our models and the
models by \citet{leung2020a} agree in the dependency of the respective isotope
production on metallicity. However, a discrepancy is visible in the yields of
selected isotopes. $^{30}$Si and $^{34}$S are produced less in our simulations,
showing differences of one order of magnitude in some cases.  Smaller
differences are apparent for $^{44}$Ti (yields by \citet{leung2020a} are
higher), $^{55}$Mn (yields by \citet{leung2020a} are lower) and $^{64}$Zn
(yields by \citet{leung2020a} are lower). In all three cases our models give
better matches relative to solar values since about solar values of Mn are
reached and the under-/overproduction of Zn/Ti is decreased.
Figure~\ref{fig:ni56-57} illustrates the abundances of $^{57}$Ni over $^{56}$Ni
as discussed in Section~\ref{sec:ni}. For comparison three models of
\citet{leung2020a} are included (at zero, $0.1$, $1$, and $5\,Z_\odot$, dashed
lines). They represent models with a total mass of $1.0\,M_\odot$ and
$1.1\,M_\odot$ and follow the same trend as our models, both regarding
metallicity and $^{56}$Ni mass. However, the models of \citet{leung2020a} show
higher values of $^{57}$Ni.  The differences in the nucleosynthesis yields of
our models compared to \citet{leung2020a} are attributed to the different
numerical treatments. The details of the approach used for the hydrodynamic
simulations of our models are given in \citet{gronow2020a, gronow2021a} and are
more accurate for the multi-dimensional structure of the double detonation.
The size of the nuclear reaction network is critical to capture the energetics
accurately as pointed out earlier.

\section{Influence on Galactic Chemical Evolution}
\label{sec:gce}
In this section, we incorporate the metallicity-dependent sub-M$_\mathrm{Ch}$
WD explosion yields presented in Section~\ref{sec:results} in the GCE code
\texttt{OMEGA+} \citep{cote2018a} in order to explore their impact on the
predicted evolution of Mn in the solar neighbourhood (see also
\citealt{lach2020a}). \texttt{OMEGA+} is a two-zone model that consists of a
central galaxy surrounded by a large gas reservoir that fills the host dark
matter halo. In addition to the star formation and chemical enrichment
processes, the code includes parametrized large-scale gas circulation processes
such as galactic inflows and outflows.  For this work, the GCE code is
calibrated to broadly reproduce general properties of the Milky Way, such as
the current star formation rate, gas inflow rate, stellar-to-gas mass ratio,
and CC and Type Ia SN rates\footnote[2]{\url{
https://github.com/becot85/JINAPyCEE/blob/master/DOC/OMEGA\%2B_Milky_Way_model.ipynb}}.

Although \texttt{OMEGA+} can account for a wide range of astrophysical sites,
we only include in this work the contribution of CC~SNe, low- and
intermediate-mass (LIMS) stars, and SNe~Ia. The nucleosynthesis ejecta are
assumed to mix homogeneously within the galactic gas, but the lifetime of stars
and the delay-time distribution (DTD) of SNe Ia are taken into account in the
enrichment process (see \citealt{ritter2018a} for more details). Given the
exploratory nature of our GCE calculations, we assume that all SNe Ia are
sub-M$_\mathrm{Ch}$ explosions, with the goal to explore how such a SN Ia
channel can contribute to the production of Mn in our Galaxy. As a test case,
we adopt the metallicity-dependent yields for the model with an initial core of
1\,M$_\odot$ and a shell mass of 0.03\,M$_\odot$, Model M10\_03. Although the yields
are metallicity-dependent, we adopt the same DTD for all metallicities, which
comes from the double detonation SN Ia prescription with a He shell found in
\cite{ruiter2014a}. This DTD is normalized such that about 10$^{-3}$ SNe~Ia
occur in total per unit of stellar mass formed, which is sufficient to account
for the overall SN Ia rate observed in nearby galaxies (see Table~5 in
\citealt{cote2016a}).

In all of our GCE calculations, the mass- and metallicity-dependent yields for
LIMS stars are taken from \cite{cristallo2015a}. For massive stars, we consider
both the mass- and metallicity-dependent yields of \citetalias{limongi2018a}
and \citetalias{nomoto2013a}. For \citetalias{limongi2018a}, we adopt the
mixture of rotation velocities introduced and described in
\cite{prantzos2018a}, which depends on metallicity. These yields are applied to
all massive stars from 8 to 100\,M$_\odot$. For \citetalias{nomoto2013a}, we
assume that 50\,\% of the stars above an initial mass of 20\,M$_\odot$ end up
as hypernovae, and apply the yields to all massive stars from 8 to
50\,M$_\odot$. This lower maximum threshold mass relative to the
\citetalias{limongi2018a} case was chosen in order to avoid overproducing the
amount of metals synthesized by massive stars.

\begin{figure}
\centering
    \includegraphics[width=3.3in]{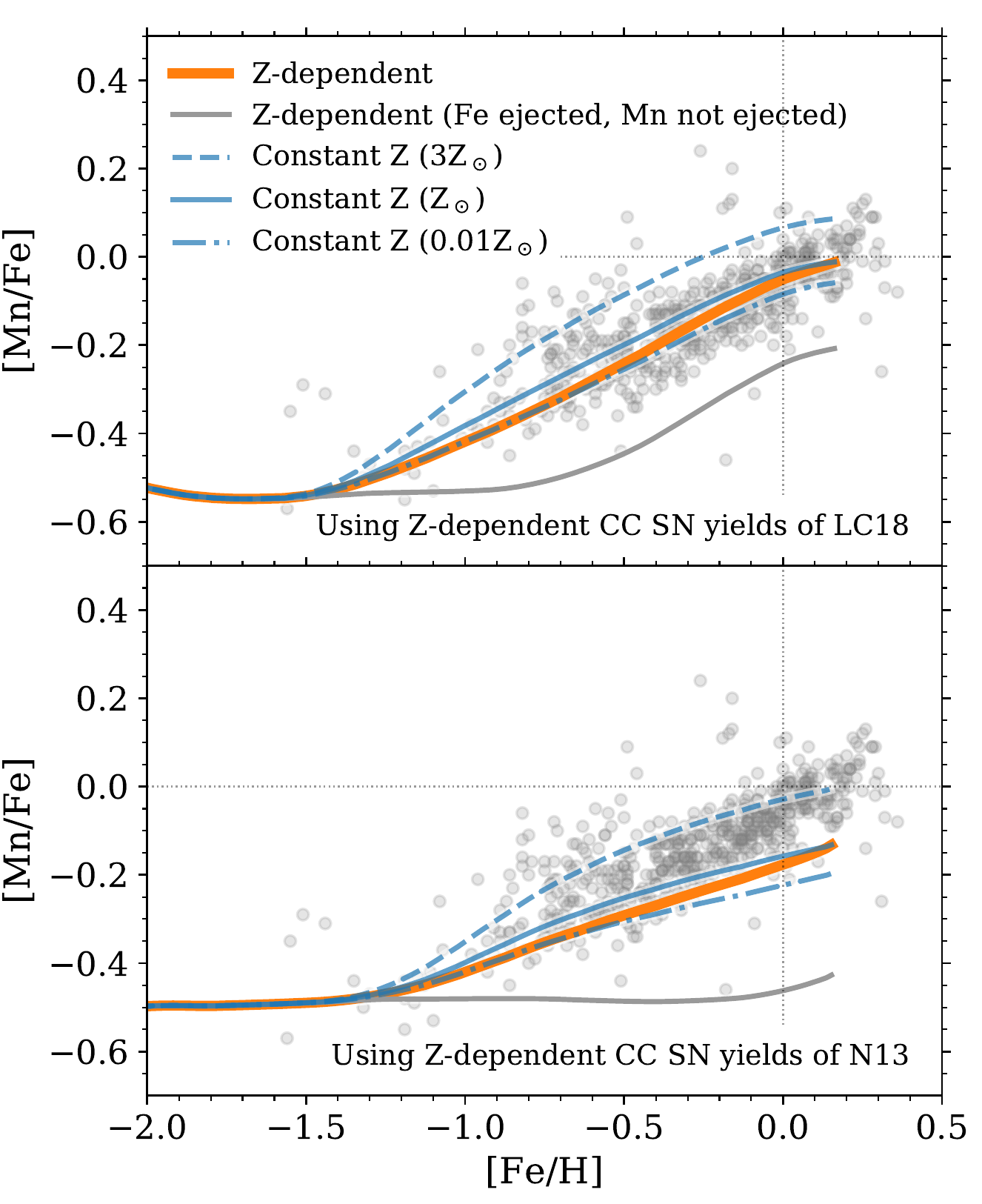}
    \caption{Chemical evolution of Mn in the solar neighbourhood, as predicted
        by \texttt{OMEGA+} (lines, Section~\ref{sec:gce}) and derived from
        stellar spectroscopy (dots, \citealt{battistini2015a}). All predictions
        include the contribution of Type Ia and CC SNe, using
        the massive star yields of \citet[LC18]{limongi2018a} and
        \citet[N13]{nomoto2013a} for the top and bottom panels, respectively.
        In each panel, the three blue lines show our predictions when we assume
        that all SNe Ia eject the yields of a given metallicity. The orange
        line shows our prediction when adopting our metallicity-dependent SN~Ia
        yields. The grey line shows the metallicity-dependent case, but when
        SNe~Ia are assumed to only eject Fe, without Mn, while CC~SNe still
        eject both Fe and Mn.}
        \label{fig:gce_Mn}
\end{figure}

Figure~\ref{fig:gce_Mn} shows the predicted evolution of [Mn/Fe] as a function
of [Fe/H] in the solar neighbourhood, assuming different CC SN yields and
different SN Ia treatments. The required contribution of SNe~Ia in order to
reproduce the upward Mn trend depends on the adopted CC SN yields (see also
\citealt{lach2020a}). As can be seen from the grey solid line, the massive star
yields of \citetalias{limongi2018a} contribute to the rise of Mn at
[Fe/H]~$>-0.5$ (upper panel), as opposed to the yields of
\citetalias{nomoto2013a} (lower panel).  In fact, when using our
metallicity-dependent SN~Ia yields (orange line) along with the massive star
yields of \citetalias{limongi2018a}, our prediction agrees relatively well with
the spectroscopic data \citep{battistini2015a}. We point out that the CC SN
yields used in \citet{seitenzahl2013b} are similar to \citetalias{nomoto2013a}.
Accordingly the chemical evolution of Mn presented in Figure 1 of
\citet{seitenzahl2013b} (thick blue line) is similar to the one in the bottom
panel of Figure~\ref{fig:gce_Mn} (orange line).

To visualize the impact of using metallicity-dependent SN~Ia yields, we ran
three additional GCE calculations in which we assumed that all SNe~Ia eject the
yields of a specific metallicity (0.01\,$Z_\odot$, \,$Z_\odot$, and
3\,$Z_\odot$, see blue lines in Figure~\ref{fig:gce_Mn}). As expected, using
higher-metallicity yields lead to higher [Mn/Fe] ratios. Although our
metallicity-dependent predictions are similar to the ones with a constant
metallicity of 0.01\,$Z_\odot$ and \,$Z_\odot$, the effect of metallicity is
still important for extending and maintaining the upward [Mn/Fe] trend at
[Fe/H]~$>0$.

\begin{figure*}
    \centering
    \includegraphics[width=6.8in]{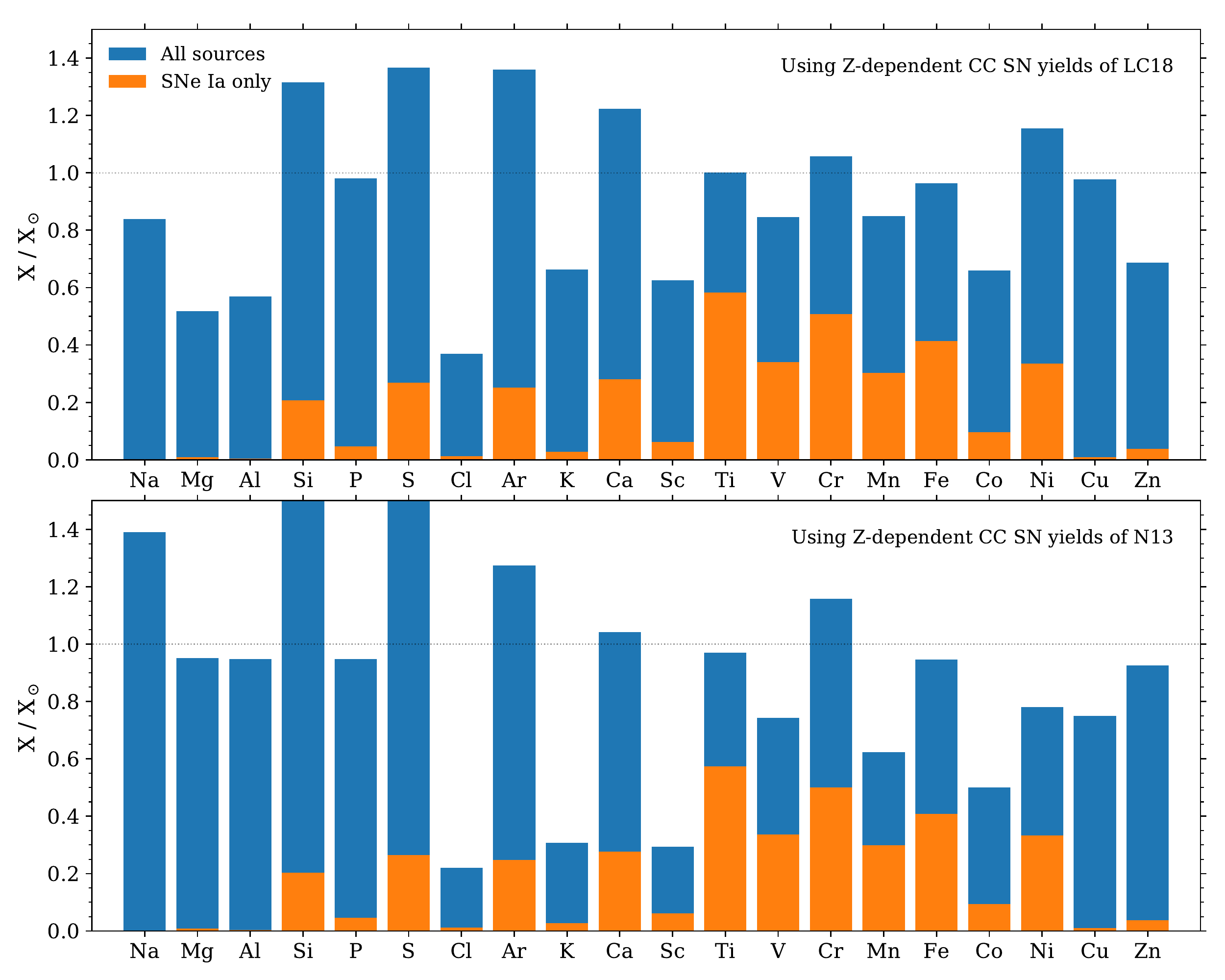}
    \caption{Predicted solar elemental distribution normalized to the solar
        abundances of \cite{asplund2009a}. The blue bars show our chemical
        evolution prediction when combining the contribution of CC SNe
        (\textit{top:} \citealt{limongi2018a}, \textit{bottom:}
        \citealt{nomoto2013a}), low- and intermediate-mass stars
    \citep{cristallo2015a}, and SNe Ia (using the metallicity-dependent yields
from this work). The orange bands show the specific contrition of SNe Ia within
the combined predicted abundances.}
    \label{fig:gce_solar}
\end{figure*}

In Figure~\ref{fig:gce_solar}, we show our predicted solar elemental abundances
when using our metallicity-dependent SN~Ia yields along with the
\citetalias{limongi2018a} yields (top) and \citetalias{nomoto2013a} yields
(bottom).  As described above, using the \citetalias{limongi2018a} yields
suggests that Mn could be significantly produced by massive stars, which would
reduce the need for SNe~Ia originating from explosions of M$_\mathrm{Ch}$ WDs.
In fact, when assuming that sub-M$_\mathrm{Ch}$ double detonations are the
dominant SN~Ia channel, our GCE model can account for more than 80\,\% of the
solar Mn abundance. This prediction, however, is affected by several
uncertainties including the exact number of SNe~Ia that occurred in the Milky
Way prior to the formation of the Solar System. Furthermore, the use of other
CC SNe yields still requires some contribution of M$_\mathrm{Ch}$ WD
explosions. The bottom panel in Figure~\ref{fig:gce_solar} illustrates this as
the Mn production is much lower using \citetalias{nomoto2013a} yields compared
with \citetalias{limongi2018a} yields.

Given the nucleosynthesis signature of our sub-M$_\mathrm{Ch}$ models, the
maximum contribution of these SNe~Ia to the evolution of Mn is limited by other
elements. In particular, Ti and Cr are produced in a substantial amount in the
ejecta of the adopted SN~Ia models, and as shown in Figure~\ref{fig:gce_solar},
our GCE prediction already fills the solar composition for these elements.
Adjusting the contribution of SNe~Ia to reproduce 100\% of the solar Mn
abundance would lead to an overproduction of Ti and Cr. Ca and Ni are
overestimated in our GCE calculations when using the massive star yields of
\citetalias{limongi2018a}, but it is not the case when using the
\citetalias{nomoto2013a} yields.

\section{Conclusions}
\label{sec:summary}
Our parameter study followed up on previous work by \citet{gronow2021a}.
Explosions of sub-M$_{\text{Ch}}$ WDs with varying core and shell masses were
simulated at different metallicities. The core mass lay between
$0.8$\,$M_\odot$ and $1.1$\,$M_\odot$ and the initial shell mass was in the
range $0.02$\,$M_\odot$ to $0.1$\,$M_\odot$. Models at solar metallicity of the
zero-age main sequence progenitor have been presented by \citet{gronow2021a}.
We now included metallicities of $0.01$, $0.1$, and $3$\,$Z_\odot$ and showed
that the results of the postprocessing step at these metallicities are in good
agreement with a full re-calculation of the hydrodynamic model allowing to
decrease the computational costs of our study.

We found that the influence of the metallicity varies with the model. The
impact is larger on the nucleosynthetic yields of the core detonation than on
those of the He detonation. In accordance with this, the nucleosynthetic yields
produced in the He detonation do not show significant changes with varying
metallicity up to elements as heavy as $^{44}$Ti. However, the amount of
$^{55}$Mn produced in the He detonation increases by one order of magnitude
with each metallicity increase (from $0.01\,Z_\odot$ to $0.1\,Z_\odot$ to
$1\,Z_\odot$ to $3\,Z_\odot$) since the presence of $^{14}$N and $^{22}$Ne
(mixed into the shell during the relaxation) supports its production (see
Section~\ref{sec:manganese}). The production of $^{54}$Fe and $^{58}$Ni
increases with metallicity as well. This is caused by a shift in the leading
reactions toward neutron-rich isotopes.

This neutron-excess also results in a change of the nucleosynthetic yields
produced in the core detonation. For $Z= 3Z_\odot$, $^{54}$Fe and $^{58}$Ni
increase to four times the value obtained at $0.01$\,$Z_\odot$. The affect is
particularly strong on $^{55}$Mn \citep{seitenzahl2013b} causing a significant
increase with metallicity.

The models are further analysed regarding changes in the elemental ratios
relative to Fe compared to solar values. All models show similar features. Most
prominent, an odd-even effect is visible in the production of IMEs.  This is
due to the higher stability of elements with an even atomic number. In
addition, the production of elements with an odd atomic number depends on the
metallicity of the white dwarf (increasing with metallicity). The models also
confirm that Cu and Zn are in large part produced in He
detonations.  Nevertheless, double detonations have sub-solar Cu and Zn
to Fe ratios due to the sub-solar production of Co, Cu, and Zn in the core
detonation. This is a feature of a pure detonation \citep{lach2020a} and
suppresses a rise of the ratio to solar values. The contribution of the core
detonation to the ratio of the double detonation is proportionally larger than
the one of the He detonation. The super-solar production of $^{44}$Ti is
caused by the He detonation. However, we do point out that some models
reproduce solar values.  While the ratios of Cu and Zn to Fe are higher
for models with larger He shells, models with lower He shell masses show a
better match with observations. Both factors need to be taken into account when
a model of a sub-M$_{\text{Ch}}$ WD is varified as progenitor of a SN\,Ia.

Only few parameter studies involving different metallicities of a
sub-M$_{\text{Ch}}$ WD were carried out \citep{shigeyama1992a, sim2010a,
shen2018b, leung2020a}. Among them only \citet{leung2020a} analyze explosions
of WDs with a He shell. We nevertheless see the same relative change in the
nucleosynthetic yields produced in the core detonation and the pure detonations
of sub-M$_{\text{Ch}}$ WDs by \citet{shigeyama1992a}, \citet{sim2010a}, and
\citet{shen2018b}. The relation between metallicity and $^{56}$Ni production
found by \citet{timmes2003a} and \citet{shen2018b} is confirmed.  The
dependence is not as strong in our models as the increase of the metallicity
only leads to a decrease of about 13\% on average. It is, however, in agreement
with a change of 15\% found by \citet{ohkubo2006a} who consider metallicities
between $0.001$\,$Z_\odot$ and $0.05$\,$Z_\odot$.

The initial mass configurations of some models in \citet{leung2020a} are in
good agreement with our M10\_10 models. The nucleosynthetic yields of some
isotopes show differences. For example, the $^{56}$Ni production is higher in
our models and the $^{30}$Si and $^{34}$S production lower. However, the trends
in the production of the isotopes are the same depending on metallicity. The
differences in the detailed nucleosynthetic yields are presumably caused by the
different numerical treatments and initial setup.

Our parameter study shows that super-solar values of Mn can be reached in
double detonations of sub-M$_{\text{Ch}}$ WDs \citep[see also][]{lach2020a}.
If such a channel is assumed to be the dominant source of SNe Ia, explosions of
sub-M$_{\text{Ch}}$ WDs could contribute significantly to the increasing
[Mn/Fe] trend at [Fe/H] $> -1$ observed in the solar neighborhood.
\citet{seitenzahl2013b} find that a contribution of at least 50\,\% by
M$_{\text{Ch}}$ WD to SNe Ia is necessary to reach solar values of [Mn/Fe]
\citep[see also][]{eitner2020a,kobayashi2019b}. This requirement is weakened by
our analysis indicating that a larger fraction of SNe Ia originates from
sub-M$_{\text{Ch}}$ WDs if CC SN yields of \citetalias{limongi2018a} are
considered.  When adopting the massive-star yields of
\citetalias{limongi2018a}, along with our metallicity-dependent SN Ia yields,
our GCE predictions can account for more than 80\% of the solar Mn, without the
contribution of M$_{\text{Ch}}$ WD explosions. It is important
that the contribution of the He shell detonation is included in the models. The
Mn production during this detonation significantly raises the total Mn over Fe
ratio. Our GCE results are obtained from a rather simplified galaxy model, and
should therefore be taken as exploratory.

\begin{acknowledgements} We thank Shing-Chi Leung for providing details
	on the treatment of metallicity effects in his simulations. This work
	was supported by the Deutsche Forschungsgemeinschaft (DFG, German
	Research Foundation) -- Project-ID 138713538 -- SFB 881 (`The Milky Way
	System'', subproject A10).  SG, FL, and FKR acknowledge support by the
	Klaus Tschira Foundation. BC was supported by the ERC Consolidator
	Grant (Hungary) funding scheme (Project RADIOSTAR, G.A. n. 724560) and
	by the Lend\"ulet grant (LP17- 2014) of the Hungarian Academy of
	Sciences.  IRS was supported by the Australian Research Council through
	grant number FT160100028. This article is based upon work from the
	“ChETEC” COST Action (CA16117), supported by COST (European Cooperation
	in Science and Technology).  CEC acknowledges support by the European
	Research Council (ERC) under the European Union's Horizon 2020 research
	and innovation programme under grant agreement No. 759253. NumPy and
	SciPy \citep{oliphant2007a}, IPython \citep{perez2007a}, and Matplotlib
	\citep{hunter2007a} were used for data processing and plotting. The
	authors gratefully acknowledge the Gauss Centre for Supercomputing e.V.
	(www.gauss-centre.eu) for funding this project by providing computing
	time on the GCS Supercomputer JUWELS \citep{juwels2019} at J\"{u}lich
	Supercomputing Centre (JSC). This work was performed using the
	Cambridge Service for Data Driven Discovery (CSD3), part of which is
	operated by the University of Cambridge Research Computing on behalf of
	the STFC DiRAC HPC Facility (www.dirac.ac.uk).  The DiRAC component of
	CSD3 was funded by BEIS capital funding via STFC capital grants
	ST/P002307/1 and ST/R002452/1 and STFC operations grant ST/R00689X/1.
	DiRAC is part of the National e-Infrastructure.  This research was
	undertaken with the assistance of resources from the National
Computational Infrastructure (NCI Australia), an NCRIS enabled capability
supported by the Australian Government.
\end{acknowledgements}

\bibliographystyle{aa}
\bibliography{astrofritz}{}

\begin{appendix}
\section{Abundances tables}
\label{sec:aa}
We list the nucleosynthesis yields of our models in Tables \ref{app:stab0803_1}
to \ref{app:rad1105_1}. They are created in the same way as in
\citet{gronow2021a}. The yields for the solar metallicity model are already
included in \citet{gronow2021a}. The stable nuclides, radioactive nuclides with
lifetimes lower than 2\,Gyr decayed to stability and radioactive nuclides with
longer lifetimes at time $t=100\,\mathrm{s}$ are listed in Table
\ref{app:stab0803_1} to \ref{app:stab1105_1}. Nucleosynthesis yields of
selected radioactive nuclides at $t=100\,\mathrm{s}$ are given in Table
\ref{app:rad0803_1} to \ref{app:rad1105_1}.

    \small
\begin{table*}
\centering
\caption{Asymptotic nucleosynthesis yields (in solar masses) for Model M08\_03 with 0.01, 0.1, and 3$\,Z_\odot$.}
\label{app:stab0803_1}
\begin{tabular}{l|cc|cc|cc}
\hline
& \multicolumn{2}{c} {\text{M08\_03\_001}} & \multicolumn{2}{c} {\text{M08\_03\_01}} & \multicolumn{2}{c} {\text{M08\_03\_3}}\\
 & He det & core det & He det & core det & He det & core det\\
 & [$M_\odot$] & [$M_\odot$] & [$M_\odot$] & [$M_\odot$] & [$M_\odot$] & [$M_\odot$]\\ \hline
$^{12}$C & 3.43e-03 & 1.26e-02 & 3.43e-03 & 1.26e-02 & 3.33e-03 & 1.21e-02\\
$^{13}$C & 1.81e-09 & 2.42e-07 & 1.80e-09 & 2.32e-07 & 1.19e-09 & 9.05e-07\\
$^{14}$N & 6.43e-07 & 9.97e-06 & 1.73e-06 & 9.67e-06 & 3.67e-05 & 1.29e-05\\
$^{15}$N & 2.33e-07 & 2.11e-08 & 2.47e-07 & 2.06e-08 & 6.49e-07 & 2.62e-08\\
$^{16}$O & 2.51e-03 & 1.44e-01 & 2.52e-03 & 1.44e-01 & 2.83e-03 & 1.46e-01\\
\hline
\end{tabular}
\end{table*}
\begin{table*}
\centering
\caption*{Table \ref{app:stab0803_1} continued.}
\label{app:stab0803_2}
\begin{tabular}{l|cc|cc|cc}
\hline
& \multicolumn{2}{c} {\text{M08\_03\_001}} & \multicolumn{2}{c} {\text{M08\_03\_01}} & \multicolumn{2}{c} {\text{M08\_03\_3}}\\
 & He det & core det & He det & core det & He det & core det\\ 
 & [$M_\odot$] & [$M_\odot$] & [$M_\odot$] & [$M_\odot$] & [$M_\odot$] & [$M_\odot$]\\ \hline
$^{50}$Cr & 1.30e-07 & 1.39e-04 & 1.30e-07 & 1.37e-04 & 1.29e-07 & 6.93e-04\\
$^{52}$Cr & 8.65e-07 & 6.20e-03 & 8.78e-07 & 6.21e-03 & 1.35e-06 & 4.87e-03\\
$^{53}$Cr & 2.90e-08 & 4.66e-04 & 3.19e-08 & 4.62e-04 & 1.55e-07 & 8.24e-04\\
$^{54}$Cr & 2.74e-10 & 2.17e-08 & 2.82e-09 & 3.62e-08 & 1.19e-07 & 1.42e-06\\
$^{55}$Mn & 1.09e-07 & 2.19e-03 & 1.28e-07 & 2.17e-03 & 8.88e-07 & 4.61e-03\\
\hline
\end{tabular}
\end{table*}
\newpage

\small
\begin{table*}
\centering
\caption{Asymptotic nucleosynthesis yields (in solar masses) for Model M08\_05 with 0.01, 0.1, and 3$\,Z_\odot$.}
\label{app:stab0805_1}
\begin{tabular}{l|cc|cc|cc}
\hline
& \multicolumn{2}{c} {\text{M08\_05\_001}} & \multicolumn{2}{c} {\text{M08\_05\_01}} & \multicolumn{2}{c} {\text{M08\_05\_3}}\\
 & He det & core det & He det & core det & He det & core det\\
 & [$M_\odot$] & [$M_\odot$] & [$M_\odot$] & [$M_\odot$] & [$M_\odot$] & [$M_\odot$]\\ \hline
$^{12}$C & 2.34e-03 & 7.55e-03 & 2.33e-03 & 7.55e-03 & 2.27e-03 & 7.25e-03\\
$^{13}$C & 5.16e-10 & 7.76e-08 & 5.15e-10 & 7.40e-08 & 3.61e-10 & 4.36e-07\\
$^{14}$N & 3.44e-07 & 6.82e-06 & 1.42e-06 & 6.57e-06 & 3.62e-05 & 1.08e-05\\
$^{15}$N & 3.46e-08 & 1.49e-08 & 3.62e-08 & 1.45e-08 & 9.03e-08 & 1.95e-08\\
$^{16}$O & 6.07e-03 & 1.16e-01 & 6.10e-03 & 1.16e-01 & 6.61e-03 & 1.18e-01\\
\hline
\end{tabular}
\end{table*}
\begin{table*}
\centering
\caption*{Table \ref{app:stab0805_1} continued.}
\label{app:stab0805_2}
\begin{tabular}{l|cc|cc|cc}
\hline
& \multicolumn{2}{c} {\text{M08\_05\_001}} & \multicolumn{2}{c} {\text{M08\_05\_01}} & \multicolumn{2}{c} {\text{M08\_05\_3}}\\
 & He det & core det & He det & core det & He det & core det\\ 
 & [$M_\odot$] & [$M_\odot$] & [$M_\odot$] & [$M_\odot$] & [$M_\odot$] & [$M_\odot$]\\ \hline
$^{50}$Cr & 7.60e-06 & 1.33e-04 & 7.69e-06 & 1.31e-04 & 1.12e-05 & 6.63e-04\\
$^{52}$Cr & 8.92e-04 & 7.37e-03 & 8.84e-04 & 7.39e-03 & 6.40e-04 & 5.77e-03\\
$^{53}$Cr & 2.00e-05 & 5.39e-04 & 2.01e-05 & 5.35e-04 & 2.20e-05 & 9.48e-04\\
$^{54}$Cr & 2.18e-10 & 1.96e-08 & 1.52e-09 & 2.51e-08 & 6.56e-08 & 1.19e-06\\
$^{55}$Mn & 3.04e-05 & 2.65e-03 & 3.04e-05 & 2.63e-03 & 2.76e-05 & 5.77e-03\\
\hline
\end{tabular}
\end{table*}
\newpage

\small
\begin{table*}
\centering
\caption{Asymptotic nucleosynthesis yields (in solar masses) for Model M08\_10 with 0.01, 0.1, and 3$\,Z_\odot$.}
\label{app:stab0810_1}
\begin{tabular}{l|cc|cc|cc}
\hline
& \multicolumn{2}{c} {\text{M08\_10\_001}} & \multicolumn{2}{c} {\text{M08\_10\_01}} & \multicolumn{2}{c} {\text{M08\_10\_3}}\\
 & He det & core det & He det & core det & He det & core det\\
 & [$M_\odot$] & [$M_\odot$] & [$M_\odot$] & [$M_\odot$] & [$M_\odot$] & [$M_\odot$]\\ \hline
$^{12}$C & 1.18e-04 & 1.07e-03 & 1.17e-04 & 1.07e-03 & 1.17e-04 & 1.04e-03\\
$^{13}$C & 7.87e-11 & 6.23e-11 & 7.69e-11 & 5.96e-11 & 4.72e-11 & 7.91e-10\\
$^{14}$N & 1.62e-07 & 1.02e-08 & 1.22e-06 & 8.98e-09 & 3.53e-05 & 1.02e-07\\
$^{15}$N & 1.00e-08 & 1.14e-09 & 1.05e-08 & 1.36e-09 & 2.70e-08 & 4.79e-10\\
$^{16}$O & 9.09e-03 & 8.03e-02 & 9.10e-03 & 8.02e-02 & 9.60e-03 & 8.17e-02\\
\hline
\end{tabular}
\end{table*}
\begin{table*}
\centering
\caption*{Table \ref{app:stab0810_1} continued.}
\label{app:stab0810_2}
\begin{tabular}{l|cc|cc|cc}
\hline
& \multicolumn{2}{c} {\text{M08\_10\_001}} & \multicolumn{2}{c} {\text{M08\_10\_01}} & \multicolumn{2}{c} {\text{M08\_10\_3}}\\
 & He det & core det & He det & core det & He det & core det\\ 
 & [$M_\odot$] & [$M_\odot$] & [$M_\odot$] & [$M_\odot$] & [$M_\odot$] & [$M_\odot$]\\ \hline
$^{50}$Cr & 4.27e-05 & 1.21e-04 & 4.23e-05 & 1.19e-04 & 3.49e-05 & 6.06e-04\\
$^{52}$Cr & 7.36e-03 & 7.93e-03 & 7.37e-03 & 7.94e-03 & 7.90e-03 & 6.23e-03\\
$^{53}$Cr & 9.66e-05 & 5.68e-04 & 9.77e-05 & 5.65e-04 & 1.36e-04 & 9.99e-04\\
$^{54}$Cr & 3.27e-10 & 3.32e-08 & 6.48e-10 & 1.94e-07 & 9.85e-09 & 8.86e-07\\
$^{55}$Mn & 9.20e-04 & 2.83e-03 & 9.22e-04 & 2.81e-03 & 9.72e-04 & 6.29e-03\\
\hline
\end{tabular}
\end{table*}
\newpage

\small
\begin{table*}
\centering
\caption{Asymptotic nucleosynthesis yields (in solar masses) for Model M09\_03 with 0.01, 0.1, and 3$\,Z_\odot$.}
\label{app:stab0903_1}
\begin{tabular}{l|cc|cc|cc}
\hline
& \multicolumn{2}{c} {\text{M09\_03\_001}} & \multicolumn{2}{c} {\text{M09\_03\_01}} & \multicolumn{2}{c} {\text{M09\_03\_3}}\\
 & He det & core det & He det & core det & He det & core det\\
 & [$M_\odot$] & [$M_\odot$] & [$M_\odot$] & [$M_\odot$] & [$M_\odot$] & [$M_\odot$]\\ \hline
$^{12}$C & 3.52e-03 & 4.94e-03 & 3.51e-03 & 4.94e-03 & 3.42e-03 & 4.74e-03\\
$^{13}$C & 1.07e-09 & 6.50e-08 & 1.07e-09 & 6.18e-08 & 7.16e-10 & 3.37e-07\\
$^{14}$N & 3.59e-07 & 3.81e-06 & 1.42e-06 & 3.67e-06 & 3.57e-05 & 5.32e-06\\
$^{15}$N & 5.76e-08 & 8.09e-09 & 6.03e-08 & 7.90e-09 & 1.50e-07 & 1.03e-08\\
$^{16}$O & 3.78e-03 & 9.17e-02 & 3.79e-03 & 9.15e-02 & 4.17e-03 & 9.31e-02\\
\hline
\end{tabular}
\end{table*}
\begin{table*}
\centering
\caption*{Table \ref{app:stab0903_1} continued.}
\label{app:stab0903_2}
\begin{tabular}{l|cc|cc|cc}
\hline
& \multicolumn{2}{c} {\text{M09\_03\_001}} & \multicolumn{2}{c} {\text{M09\_03\_01}} & \multicolumn{2}{c} {\text{M09\_03\_3}}\\
 & He det & core det & He det & core det & He det & core det\\ 
 & [$M_\odot$] & [$M_\odot$] & [$M_\odot$] & [$M_\odot$] & [$M_\odot$] & [$M_\odot$]\\ \hline
$^{50}$Cr & 1.37e-06 & 1.41e-04 & 1.37e-06 & 1.38e-04 & 1.34e-06 & 7.01e-04\\
$^{52}$Cr & 5.26e-06 & 9.50e-03 & 5.17e-06 & 9.51e-03 & 3.14e-06 & 7.45e-03\\
$^{53}$Cr & 4.67e-07 & 6.72e-04 & 4.65e-07 & 6.67e-04 & 4.37e-07 & 1.18e-03\\
$^{54}$Cr & 3.18e-10 & 2.14e-08 & 2.99e-09 & 2.24e-08 & 1.25e-07 & 1.01e-06\\
$^{55}$Mn & 2.61e-07 & 3.37e-03 & 2.76e-07 & 3.34e-03 & 9.72e-07 & 7.50e-03\\
\hline
\end{tabular}
\end{table*}
\newpage

\small
\begin{table*}
\centering
\caption{Asymptotic nucleosynthesis yields (in solar masses) for Model M09\_05 with 0.01, 0.1, and 3$\,Z_\odot$.}
\label{app:stab0905_1}
\begin{tabular}{l|cc|cc|cc}
\hline
& \multicolumn{2}{c} {\text{M09\_05\_001}} & \multicolumn{2}{c} {\text{M09\_05\_01}} & \multicolumn{2}{c} {\text{M09\_05\_3}}\\
 & He det & core det & He det & core det & He det & core det\\
 & [$M_\odot$] & [$M_\odot$] & [$M_\odot$] & [$M_\odot$] & [$M_\odot$] & [$M_\odot$]\\ \hline
$^{12}$C & 4.37e-04 & 2.67e-03 & 4.36e-04 & 2.66e-03 & 4.29e-04 & 2.58e-03\\
$^{13}$C & 2.49e-11 & 3.02e-09 & 2.40e-11 & 2.83e-09 & 1.87e-11 & 3.17e-08\\
$^{14}$N & 1.66e-07 & 4.48e-07 & 1.23e-06 & 4.26e-07 & 3.54e-05 & 1.70e-06\\
$^{15}$N & 1.00e-08 & 2.92e-09 & 1.06e-08 & 2.93e-09 & 2.67e-08 & 4.12e-09\\
$^{16}$O & 7.12e-03 & 7.73e-02 & 7.14e-03 & 7.71e-02 & 7.61e-03 & 7.86e-02\\
\hline
\end{tabular}
\end{table*}
\begin{table*}
\centering
\caption*{Table \ref{app:stab0905_1} continued.}
\label{app:stab0905_2}
\begin{tabular}{l|cc|cc|cc}
\hline
& \multicolumn{2}{c} {\text{M09\_05\_001}} & \multicolumn{2}{c} {\text{M09\_05\_01}} & \multicolumn{2}{c} {\text{M09\_05\_3}}\\
 & He det & core det & He det & core det & He det & core det\\ 
 & [$M_\odot$] & [$M_\odot$] & [$M_\odot$] & [$M_\odot$] & [$M_\odot$] & [$M_\odot$]\\ \hline
$^{50}$Cr & 1.83e-05 & 1.25e-04 & 1.85e-05 & 1.22e-04 & 2.49e-05 & 6.20e-04\\
$^{52}$Cr & 5.09e-03 & 8.80e-03 & 5.10e-03 & 8.81e-03 & 5.27e-03 & 6.91e-03\\
$^{53}$Cr & 1.36e-04 & 6.21e-04 & 1.37e-04 & 6.17e-04 & 1.72e-04 & 1.09e-03\\
$^{54}$Cr & 2.85e-10 & 1.61e-08 & 7.55e-10 & 1.86e-08 & 1.84e-08 & 8.47e-07\\
$^{55}$Mn & 4.12e-04 & 3.10e-03 & 4.12e-04 & 3.07e-03 & 3.95e-04 & 6.94e-03\\
\hline
\end{tabular}
\end{table*}
\newpage

\small
\begin{table*}
\centering
\caption{Asymptotic nucleosynthesis yields (in solar masses) for Model M09\_10 with 0.01, 0.1, and 3$\,Z_\odot$.}
\label{app:stab0910_1}
\begin{tabular}{l|cc|cc|cc}
\hline
& \multicolumn{2}{c} {\text{M09\_10\_001}} & \multicolumn{2}{c} {\text{M09\_10\_01}} & \multicolumn{2}{c} {\text{M09\_10\_3}}\\
 & He det & core det & He det & core det & He det & core det\\
 & [$M_\odot$] & [$M_\odot$] & [$M_\odot$] & [$M_\odot$] & [$M_\odot$] & [$M_\odot$]\\ \hline
$^{12}$C & 3.82e-05 & 1.36e-04 & 3.82e-05 & 1.35e-04 & 3.80e-05 & 1.32e-04\\
$^{13}$C & 1.89e-09 & 5.87e-12 & 1.88e-09 & 5.74e-12 & 1.55e-09 & 2.21e-11\\
$^{14}$N & 2.49e-07 & 1.93e-10 & 1.31e-06 & 1.93e-10 & 3.55e-05 & 4.65e-10\\
$^{15}$N & 1.39e-08 & 3.94e-10 & 1.47e-08 & 4.25e-10 & 3.74e-08 & 2.34e-10\\
$^{16}$O & 8.36e-03 & 5.46e-02 & 8.37e-03 & 5.45e-02 & 8.82e-03 & 5.55e-02\\
\hline
\end{tabular}
\end{table*}
\begin{table*}
\centering
\caption*{Table \ref{app:stab0910_1} continued.}
\label{app:stab0910_2}
\begin{tabular}{l|cc|cc|cc}
\hline
& \multicolumn{2}{c} {\text{M09\_10\_001}} & \multicolumn{2}{c} {\text{M09\_10\_01}} & \multicolumn{2}{c} {\text{M09\_10\_3}}\\
 & He det & core det & He det & core det & He det & core det\\ 
 & [$M_\odot$] & [$M_\odot$] & [$M_\odot$] & [$M_\odot$] & [$M_\odot$] & [$M_\odot$]\\ \hline
$^{50}$Cr & 2.26e-04 & 1.07e-04 & 2.25e-04 & 1.05e-04 & 1.86e-04 & 5.33e-04\\
$^{52}$Cr & 3.98e-03 & 8.10e-03 & 3.98e-03 & 8.12e-03 & 4.08e-03 & 6.35e-03\\
$^{53}$Cr & 6.05e-05 & 5.69e-04 & 6.11e-05 & 5.66e-04 & 8.41e-05 & 9.98e-04\\
$^{54}$Cr & 2.57e-10 & 1.63e-08 & 4.38e-10 & 2.02e-08 & 7.17e-09 & 1.03e-06\\
$^{55}$Mn & 3.68e-04 & 2.89e-03 & 3.69e-04 & 2.86e-03 & 3.83e-04 & 6.49e-03\\
\hline
\end{tabular}
\end{table*}
\newpage

\small
\begin{table*}
\centering
\caption{Asymptotic nucleosynthesis yields (in solar masses) for Model M10\_02 with 0.01, 0.1, and 3$\,Z_\odot$.}
\label{app:stab1002_1}
\begin{tabular}{l|cc|cc|cc}
\hline
& \multicolumn{2}{c} {\text{M10\_02\_001}} & \multicolumn{2}{c} {\text{M10\_02\_01}} & \multicolumn{2}{c} {\text{M10\_02\_3}}\\
 & He det & core det & He det & core det & He det & core det\\
 & [$M_\odot$] & [$M_\odot$] & [$M_\odot$] & [$M_\odot$] & [$M_\odot$] & [$M_\odot$]\\ \hline
$^{12}$C & 1.69e-03 & 1.97e-03 & 1.69e-03 & 1.97e-03 & 1.65e-03 & 1.90e-03\\
$^{13}$C & 3.51e-10 & 1.77e-08 & 3.50e-10 & 1.67e-08 & 2.44e-10 & 9.78e-08\\
$^{14}$N & 2.02e-07 & 9.45e-07 & 1.25e-06 & 9.07e-07 & 3.50e-05 & 1.55e-06\\
$^{15}$N & 2.72e-08 & 2.50e-09 & 2.85e-08 & 2.46e-09 & 6.99e-08 & 3.17e-09\\
$^{16}$O & 1.77e-03 & 5.67e-02 & 1.78e-03 & 5.66e-02 & 2.02e-03 & 5.76e-02\\
\hline
\end{tabular}
\end{table*}
\begin{table*}
\centering
\caption*{Table \ref{app:stab1002_1} continued.}
\label{app:stab1002_2}
\begin{tabular}{l|cc|cc|cc}
\hline
& \multicolumn{2}{c} {\text{M10\_02\_001}} & \multicolumn{2}{c} {\text{M10\_02\_01}} & \multicolumn{2}{c} {\text{M10\_02\_3}}\\
 & He det & core det & He det & core det & He det & core det\\ 
 & [$M_\odot$] & [$M_\odot$] & [$M_\odot$] & [$M_\odot$] & [$M_\odot$] & [$M_\odot$]\\ \hline
$^{50}$Cr & 1.36e-06 & 1.20e-04 & 1.37e-06 & 1.18e-04 & 1.64e-06 & 6.02e-04\\
$^{52}$Cr & 2.92e-05 & 9.50e-03 & 2.89e-05 & 9.51e-03 & 1.98e-05 & 7.44e-03\\
$^{53}$Cr & 1.29e-06 & 6.62e-04 & 1.29e-06 & 6.57e-04 & 1.35e-06 & 1.16e-03\\
$^{54}$Cr & 1.73e-10 & 1.38e-08 & 1.59e-09 & 1.55e-08 & 6.43e-08 & 7.00e-07\\
$^{55}$Mn & 1.28e-06 & 3.34e-03 & 1.29e-06 & 3.30e-03 & 1.62e-06 & 7.51e-03\\
\hline
\end{tabular}
\end{table*}
\newpage

\small
\begin{table*}
\centering
\caption{Asymptotic nucleosynthesis yields (in solar masses) for Model M10\_03 with 0.01, 0.1, and 3$\,Z_\odot$.}
\label{app:stab1003_1}
\begin{tabular}{l|cc|cc|cc}
\hline
& \multicolumn{2}{c} {\text{M10\_03\_001}} & \multicolumn{2}{c} {\text{M10\_03\_01}} & \multicolumn{2}{c} {\text{M10\_03\_3}}\\
 & He det & core det & He det & core det & He det & core det\\
 & [$M_\odot$] & [$M_\odot$] & [$M_\odot$] & [$M_\odot$] & [$M_\odot$] & [$M_\odot$]\\ \hline
$^{12}$C & 7.71e-04 & 1.24e-03 & 7.68e-04 & 1.24e-03 & 7.55e-04 & 1.20e-03\\
$^{13}$C & 4.95e-11 & 1.12e-09 & 4.92e-11 & 1.04e-09 & 3.96e-11 & 1.19e-08\\
$^{14}$N & 1.71e-07 & 1.37e-07 & 1.22e-06 & 1.29e-07 & 3.51e-05 & 5.12e-07\\
$^{15}$N & 1.27e-08 & 1.12e-09 & 1.33e-08 & 1.12e-09 & 3.35e-08 & 1.27e-09\\
$^{16}$O & 6.60e-03 & 4.85e-02 & 6.62e-03 & 4.84e-02 & 7.09e-03 & 4.93e-02\\
\hline
\end{tabular}
\end{table*}
\begin{table*}
\centering
\caption*{Table \ref{app:stab1003_1} continued.}
\label{app:stab1003_2}
\begin{tabular}{l|cc|cc|cc}
\hline
& \multicolumn{2}{c} {\text{M10\_03\_001}} & \multicolumn{2}{c} {\text{M10\_03\_01}} & \multicolumn{2}{c} {\text{M10\_03\_3}}\\
 & He det & core det & He det & core det & He det & core det\\ 
 & [$M_\odot$] & [$M_\odot$] & [$M_\odot$] & [$M_\odot$] & [$M_\odot$] & [$M_\odot$]\\ \hline
$^{50}$Cr & 3.87e-06 & 1.08e-04 & 3.90e-06 & 1.06e-04 & 5.06e-06 & 5.40e-04\\
$^{52}$Cr & 7.32e-04 & 8.74e-03 & 7.25e-04 & 8.76e-03 & 5.14e-04 & 6.87e-03\\
$^{53}$Cr & 1.27e-05 & 6.09e-04 & 1.27e-05 & 6.05e-04 & 1.22e-05 & 1.07e-03\\
$^{54}$Cr & 1.23e-10 & 1.27e-08 & 7.42e-10 & 1.42e-08 & 2.83e-08 & 6.21e-07\\
$^{55}$Mn & 1.74e-05 & 3.07e-03 & 1.73e-05 & 3.04e-03 & 1.44e-05 & 6.94e-03\\
\hline
\end{tabular}
\end{table*}
\newpage

\small
\begin{table*}
\centering
\caption{Asymptotic nucleosynthesis yields (in solar masses) for Model M10\_05 with 0.01, 0.1, and 3$\,Z_\odot$.}
\label{app:stab1005_1}
\begin{tabular}{l|cc|cc|cc}
\hline
& \multicolumn{2}{c} {\text{M10\_05\_001}} & \multicolumn{2}{c} {\text{M10\_05\_01}} & \multicolumn{2}{c} {\text{M10\_05\_3}}\\
 & He det & core det & He det & core det & He det & core det\\
 & [$M_\odot$] & [$M_\odot$] & [$M_\odot$] & [$M_\odot$] & [$M_\odot$] & [$M_\odot$]\\ \hline
$^{12}$C & 3.96e-05 & 4.45e-04 & 3.95e-05 & 4.43e-04 & 3.97e-05 & 4.30e-04\\
$^{13}$C & 1.91e-10 & 1.25e-10 & 1.88e-10 & 1.20e-10 & 9.33e-11 & 7.98e-10\\
$^{14}$N & 1.49e-07 & 2.49e-08 & 1.21e-06 & 2.40e-08 & 3.55e-05 & 7.19e-08\\
$^{15}$N & 4.65e-09 & 5.90e-10 & 4.93e-09 & 1.08e-09 & 1.25e-08 & 7.36e-10\\
$^{16}$O & 9.17e-03 & 6.04e-02 & 9.18e-03 & 6.03e-02 & 9.67e-03 & 6.15e-02\\
\hline
\end{tabular}
\end{table*}
\begin{table*}
\centering
\caption*{Table \ref{app:stab1005_1} continued.}
\label{app:stab1005_2}
\begin{tabular}{l|cc|cc|cc}
\hline
& \multicolumn{2}{c} {\text{M10\_05\_001}} & \multicolumn{2}{c} {\text{M10\_05\_01}} & \multicolumn{2}{c} {\text{M10\_05\_3}}\\
 & He det & core det & He det & core det & He det & core det\\ 
 & [$M_\odot$] & [$M_\odot$] & [$M_\odot$] & [$M_\odot$] & [$M_\odot$] & [$M_\odot$]\\ \hline
$^{50}$Cr & 2.56e-05 & 1.11e-04 & 2.54e-05 & 1.08e-04 & 2.13e-05 & 5.51e-04\\
$^{52}$Cr & 4.02e-03 & 8.43e-03 & 4.03e-03 & 8.44e-03 & 4.37e-03 & 6.61e-03\\
$^{53}$Cr & 5.77e-05 & 5.91e-04 & 5.83e-05 & 5.87e-04 & 7.91e-05 & 1.04e-03\\
$^{54}$Cr & 2.97e-10 & 1.39e-08 & 4.68e-10 & 1.69e-08 & 6.75e-09 & 7.03e-07\\
$^{55}$Mn & 4.78e-04 & 2.98e-03 & 4.79e-04 & 2.95e-03 & 4.94e-04 & 6.71e-03\\
\hline
\end{tabular}
\end{table*}
\newpage

\small
\begin{table*}
\centering
\caption{Asymptotic nucleosynthesis yields (in solar masses) for Model M10\_10 with 0.01, 0.1, and 3$\,Z_\odot$.}
\label{app:stab1010_1}
\begin{tabular}{l|cc|cc|cc}
\hline
& \multicolumn{2}{c} {\text{M10\_10\_001}} & \multicolumn{2}{c} {\text{M10\_10\_01}} & \multicolumn{2}{c} {\text{M10\_10\_3}}\\
 & He det & core det & He det & core det & He det & core det\\
 & [$M_\odot$] & [$M_\odot$] & [$M_\odot$] & [$M_\odot$] & [$M_\odot$] & [$M_\odot$]\\ \hline
$^{12}$C & 7.39e-06 & 2.46e-06 & 7.25e-06 & 2.47e-06 & 7.52e-06 & 1.12e-06\\
$^{13}$C & 4.23e-09 & 2.49e-12 & 4.21e-09 & 2.57e-12 & 3.65e-09 & 1.12e-12\\
$^{14}$N & 1.25e-07 & 3.04e-12 & 1.16e-06 & 3.13e-12 & 3.46e-05 & 1.88e-11\\
$^{15}$N & 3.15e-10 & 2.12e-12 & 3.27e-10 & 2.13e-12 & 8.11e-10 & 4.36e-11\\
$^{16}$O & 3.02e-03 & 2.70e-03 & 3.02e-03 & 2.70e-03 & 3.24e-03 & 2.75e-03\\
\hline
\end{tabular}
\end{table*}
\begin{table*}
\centering
\caption*{Table \ref{app:stab1010_1} continued.}
\label{app:stab1010_2}
\begin{tabular}{l|cc|cc|cc}
\hline
& \multicolumn{2}{c} {\text{M10\_10\_001}} & \multicolumn{2}{c} {\text{M10\_10\_01}} & \multicolumn{2}{c} {\text{M10\_10\_3}}\\
 & He det & core det & He det & core det & He det & core det\\ 
 & [$M_\odot$] & [$M_\odot$] & [$M_\odot$] & [$M_\odot$] & [$M_\odot$] & [$M_\odot$]\\ \hline
$^{50}$Cr & 1.03e-04 & 7.59e-05 & 1.03e-04 & 7.44e-05 & 8.37e-05 & 3.89e-04\\
$^{52}$Cr & 1.97e-03 & 9.27e-03 & 1.97e-03 & 9.28e-03 & 2.06e-03 & 7.22e-03\\
$^{53}$Cr & 5.50e-05 & 6.23e-04 & 5.53e-05 & 6.19e-04 & 6.65e-05 & 1.07e-03\\
$^{54}$Cr & 1.74e-09 & 1.14e-08 & 1.95e-09 & 1.72e-09 & 1.44e-08 & 2.51e-07\\
$^{55}$Mn & 2.52e-04 & 3.16e-03 & 2.53e-04 & 3.13e-03 & 2.78e-04 & 7.24e-03\\
\hline
\end{tabular}
\end{table*}
\newpage

\small
\begin{table*}
\centering
\caption{Asymptotic nucleosynthesis yields (in solar masses) for Model M11\_05 with 0.01, 0.1, and 3$\,Z_\odot$.}
\label{app:stab1105_1}
\begin{tabular}{l|cc|cc|cc}
\hline
& \multicolumn{2}{c} {\text{M11\_05\_001}} & \multicolumn{2}{c} {\text{M11\_05\_01}} & \multicolumn{2}{c} {\text{M11\_05\_3}}\\
 & He det & core det & He det & core det & He det & core det\\
 & [$M_\odot$] & [$M_\odot$] & [$M_\odot$] & [$M_\odot$] & [$M_\odot$] & [$M_\odot$]\\ \hline
$^{12}$C & 5.24e-06 & 3.11e-06 & 5.27e-06 & 3.12e-06 & 5.65e-06 & 1.23e-06\\
$^{13}$C & 1.01e-09 & 3.17e-12 & 1.00e-09 & 3.27e-12 & 8.18e-10 & 1.04e-12\\
$^{14}$N & 1.38e-07 & 7.05e-13 & 1.20e-06 & 7.21e-13 & 3.54e-05 & 4.99e-11\\
$^{15}$N & 2.43e-09 & 1.14e-12 & 2.56e-09 & 1.15e-12 & 6.49e-09 & 1.37e-10\\
$^{16}$O & 3.74e-03 & 7.48e-04 & 3.74e-03 & 7.47e-04 & 4.02e-03 & 7.62e-04\\
\hline
\end{tabular}
\end{table*}
\begin{table*}
\centering
\caption*{Table \ref{app:stab1105_1} continued.}
\label{app:stab1105_2}
\begin{tabular}{l|cc|cc|cc}
\hline
& \multicolumn{2}{c} {\text{M11\_05\_001}} & \multicolumn{2}{c} {\text{M11\_05\_01}} & \multicolumn{2}{c} {\text{M11\_05\_3}}\\
 & He det & core det & He det & core det & He det & core det\\ 
 & [$M_\odot$] & [$M_\odot$] & [$M_\odot$] & [$M_\odot$] & [$M_\odot$] & [$M_\odot$]\\ \hline
$^{50}$Cr & 4.09e-05 & 5.39e-05 & 4.10e-05 & 5.29e-05 & 4.57e-05 & 2.66e-04\\
$^{52}$Cr & 2.04e-03 & 7.80e-03 & 2.05e-03 & 7.81e-03 & 2.24e-03 & 6.11e-03\\
$^{53}$Cr & 7.19e-05 & 5.16e-04 & 7.21e-05 & 5.13e-04 & 8.16e-05 & 8.97e-04\\
$^{54}$Cr & 4.45e-09 & 7.59e-10 & 4.77e-09 & 1.14e-09 & 2.32e-08 & 1.71e-07\\
$^{55}$Mn & 2.14e-04 & 2.63e-03 & 2.15e-04 & 2.61e-03 & 2.32e-04 & 6.14e-03\\
\hline
\end{tabular}
\end{table*}
\newpage

\begin{table*}
\centering
\caption{Nucleosynthesis yields (in solar masses) for select radioactive nuclides of Model M08\_03 with 0.01, 0.1, and 3$\,Z_\odot$.}
\label{app:rad0803_1}
\begin{tabular}{l|cc|cc|cc}
\hline
& \multicolumn{2}{c} {\text{M08\_03\_001}} & \multicolumn{2}{c} {\text{M08\_03\_01}} & \multicolumn{2}{c} {\text{M08\_03\_3}}\\
 & He det & core det & He det & core det & He det & core det\\
 & [$M_\odot$] & [$M_\odot$] & [$M_\odot$] & [$M_\odot$] & [$M_\odot$] & [$M_\odot$]\\ \hline
$^{14}$C & 7.29e-08 & 4.85e-06 & 7.43e-08 & 4.62e-06 & 9.15e-08 & 8.76e-06\\
$^{22}$Na & 5.45e-07 & 2.50e-08 & 5.75e-07 & 2.50e-08 & 1.44e-06 & 1.74e-08\\
$^{26}$Al & 1.13e-05 & 1.37e-05 & 1.14e-05 & 1.37e-05 & 1.53e-05 & 5.47e-06\\
$^{32}$Si & 8.10e-10 & 7.80e-10 & 1.39e-09 & 1.05e-09 & 2.34e-09 & 8.25e-08\\
$^{32}$P & 7.78e-09 & 2.23e-07 & 8.35e-09 & 2.19e-07 & 1.27e-08 & 3.92e-06\\
\hline
\end{tabular}
\end{table*}
\newpage
\begin{table*}
\centering
\caption{Nucleosynthesis yields (in solar masses) for select radioactive nuclides of Model M08\_05 with 0.01, 0.1, and 3$\,Z_\odot$.}
\label{app:rad0805_1}
\begin{tabular}{l|cc|cc|cc}
\hline
& \multicolumn{2}{c} {\text{M08\_05\_001}} & \multicolumn{2}{c} {\text{M08\_05\_01}} & \multicolumn{2}{c} {\text{M08\_05\_3}}\\
 & He det & core det & He det & core det & He det & core det\\
 & [$M_\odot$] & [$M_\odot$] & [$M_\odot$] & [$M_\odot$] & [$M_\odot$] & [$M_\odot$]\\ \hline
$^{14}$C & 2.09e-08 & 3.37e-06 & 2.12e-08 & 3.19e-06 & 2.26e-08 & 7.51e-06\\
$^{22}$Na & 5.09e-08 & 2.15e-08 & 5.29e-08 & 2.15e-08 & 1.16e-07 & 1.47e-08\\
$^{26}$Al & 9.81e-06 & 1.12e-05 & 9.79e-06 & 1.12e-05 & 1.03e-05 & 4.40e-06\\
$^{32}$Si & 9.42e-11 & 8.05e-10 & 1.77e-10 & 1.02e-09 & 3.25e-10 & 8.08e-08\\
$^{32}$P & 1.44e-08 & 1.87e-07 & 1.47e-08 & 1.83e-07 & 1.65e-08 & 3.20e-06\\
\hline
\end{tabular}
\end{table*}
\newpage
\begin{table*}
\centering
\caption{Nucleosynthesis yields (in solar masses) for select radioactive nuclides of Model M08\_10 with 0.01, 0.1, and 3$\,Z_\odot$.}
\label{app:raad0810_1}
\begin{tabular}{l|cc|cc|cc}
\hline
& \multicolumn{2}{c} {\text{M08\_10\_001}} & \multicolumn{2}{c} {\text{M08\_10\_01}} & \multicolumn{2}{c} {\text{M08\_10\_3}}\\
 & He det & core det & He det & core det & He det & core det\\
 & [$M_\odot$] & [$M_\odot$] & [$M_\odot$] & [$M_\odot$] & [$M_\odot$] & [$M_\odot$]\\ \hline
$^{14}$C & 2.53e-12 & 2.17e-09 & 2.81e-12 & 1.29e-09 & 1.15e-11 & 5.76e-08\\
$^{22}$Na & 8.66e-09 & 1.03e-08 & 8.97e-09 & 1.04e-08 & 1.79e-08 & 7.49e-09\\
$^{26}$Al & 6.65e-07 & 6.65e-06 & 6.63e-07 & 6.67e-06 & 7.06e-07 & 2.76e-06\\
$^{32}$Si & 2.18e-12 & 8.20e-10 & 2.24e-12 & 4.33e-11 & 2.54e-12 & 1.41e-08\\
$^{32}$P & 2.00e-08 & 1.19e-07 & 2.01e-08 & 1.16e-07 & 2.14e-08 & 2.09e-06\\
\hline
\end{tabular}
\end{table*}
\newpage
\begin{table*}
\centering
\caption{Nucleosynthesis yields (in solar masses) for select radioactive nuclides of Model M09\_03 with 0.01, 0.1, and 3$\,Z_\odot$.}
\label{app:rad0903_1}
\begin{tabular}{l|cc|cc|cc}
\hline
& \multicolumn{2}{c} {\text{M09\_03\_001}} & \multicolumn{2}{c} {\text{M09\_03\_01}} & \multicolumn{2}{c} {\text{M09\_03\_3}}\\
 & He det & core det & He det & core det & He det & core det\\
 & [$M_\odot$] & [$M_\odot$] & [$M_\odot$] & [$M_\odot$] & [$M_\odot$] & [$M_\odot$]\\ \hline
$^{14}$C & 3.68e-08 & 1.93e-06 & 3.73e-08 & 1.83e-06 & 4.31e-08 & 3.71e-06\\
$^{22}$Na & 9.34e-08 & 1.38e-08 & 9.78e-08 & 1.38e-08 & 2.32e-07 & 9.36e-09\\
$^{26}$Al & 1.33e-05 & 7.04e-06 & 1.33e-05 & 7.09e-06 & 1.45e-05 & 2.74e-06\\
$^{32}$Si & 3.09e-10 & 5.35e-10 & 5.69e-10 & 6.89e-10 & 9.61e-10 & 5.47e-08\\
$^{32}$P & 1.00e-08 & 1.39e-07 & 1.05e-08 & 1.36e-07 & 1.32e-08 & 2.22e-06\\
\hline
\end{tabular}
\end{table*}
\newpage
\begin{table*}
\centering
\caption{Nucleosynthesis yields (in solar masses) for select radioactive nuclides of Model M09\_05 with 0.01, 0.1, and 3$\,Z_\odot$.}
\label{app:rad0905_1}
\begin{tabular}{l|cc|cc|cc}
\hline
& \multicolumn{2}{c} {\text{M09\_05\_001}} & \multicolumn{2}{c} {\text{M09\_05\_01}} & \multicolumn{2}{c} {\text{M09\_05\_3}}\\
 & He det & core det & He det & core det & He det & core det\\
 & [$M_\odot$] & [$M_\odot$] & [$M_\odot$] & [$M_\odot$] & [$M_\odot$] & [$M_\odot$]\\ \hline
$^{14}$C & 6.91e-11 & 2.08e-07 & 6.95e-11 & 1.95e-07 & 5.21e-11 & 1.18e-06\\
$^{22}$Na & 1.49e-08 & 1.41e-08 & 1.54e-08 & 1.41e-08 & 2.93e-08 & 1.00e-08\\
$^{26}$Al & 3.04e-06 & 7.54e-06 & 3.03e-06 & 7.58e-06 & 3.06e-06 & 2.94e-06\\
$^{32}$Si & 4.31e-12 & 2.23e-10 & 4.13e-12 & 2.40e-10 & 4.67e-12 & 3.72e-08\\
$^{32}$P & 1.54e-08 & 1.19e-07 & 1.54e-08 & 1.16e-07 & 1.63e-08 & 2.06e-06\\
\hline
\end{tabular}
\end{table*}
\newpage
\begin{table*}
\centering
\caption{Nucleosynthesis yields (in solar masses) for select radioactive nuclides of Model M09\_10 with 0.01, 0.1, and 3$\,Z_\odot$.}
\label{app:rad0910_1}
\begin{tabular}{l|cc|cc|cc}
\hline
& \multicolumn{2}{c} {\text{M09\_10\_001}} & \multicolumn{2}{c} {\text{M09\_10\_01}} & \multicolumn{2}{c} {\text{M09\_10\_3}}\\
 & He det & core det & He det & core det & He det & core det\\
 & [$M_\odot$] & [$M_\odot$] & [$M_\odot$] & [$M_\odot$] & [$M_\odot$] & [$M_\odot$]\\ \hline
$^{14}$C & 3.84e-12 & 2.80e-13 & 4.47e-12 & 1.73e-13 & 2.36e-11 & 3.61e-11\\
$^{22}$Na & 7.85e-09 & 5.73e-09 & 8.22e-09 & 5.68e-09 & 1.83e-08 & 6.08e-09\\
$^{26}$Al & 9.78e-08 & 1.84e-06 & 9.91e-08 & 1.84e-06 & 1.43e-07 & 7.96e-07\\
$^{32}$Si & 1.93e-12 & 4.35e-10 & 2.03e-12 & 7.81e-11 & 2.34e-12 & 3.41e-09\\
$^{32}$P & 1.52e-08 & 7.62e-08 & 1.55e-08 & 7.39e-08 & 1.65e-08 & 1.07e-06\\
\hline
\end{tabular}
\end{table*}
\newpage
\begin{table*}
\centering
\caption{Nucleosynthesis yields (in solar masses) for select radioactive nuclides of Model M10\_02 with 0.01, 0.1, and 3$\,Z_\odot$.}
\label{app:rad1002_1}
\begin{tabular}{l|cc|cc|cc}
\hline
& \multicolumn{2}{c} {\text{M10\_02\_001}} & \multicolumn{2}{c} {\text{M10\_02\_01}} & \multicolumn{2}{c} {\text{M10\_02\_3}}\\
 & He det & core det & He det & core det & He det & core det\\
 & [$M_\odot$] & [$M_\odot$] & [$M_\odot$] & [$M_\odot$] & [$M_\odot$] & [$M_\odot$]\\ \hline
$^{14}$C & 9.13e-09 & 4.82e-07 & 9.25e-09 & 4.55e-07 & 1.07e-08 & 1.09e-06\\
$^{22}$Na & 4.63e-08 & 7.63e-09 & 4.85e-08 & 7.62e-09 & 1.15e-07 & 5.23e-09\\
$^{26}$Al & 6.68e-06 & 4.05e-06 & 6.69e-06 & 4.08e-06 & 7.41e-06 & 1.57e-06\\
$^{32}$Si & 9.92e-11 & 2.11e-10 & 1.67e-10 & 2.53e-10 & 2.82e-10 & 2.52e-08\\
$^{32}$P & 4.98e-09 & 8.29e-08 & 5.13e-09 & 8.09e-08 & 5.94e-09 & 1.31e-06\\
\hline
\end{tabular}
\end{table*}
\newpage
\begin{table*}
\centering
\caption{Nucleosynthesis yields (in solar masses) for select radioactive nuclides of Model M10\_03 with 0.01, 0.1, and 3$\,Z_\odot$.}
\label{app:rad1003_1}
\begin{tabular}{l|cc|cc|cc}
\hline
& \multicolumn{2}{c} {\text{M10\_03\_001}} & \multicolumn{2}{c} {\text{M10\_03\_01}} & \multicolumn{2}{c} {\text{M10\_03\_3}}\\
 & He det & core det & He det & core det & He det & core det\\
 & [$M_\odot$] & [$M_\odot$] & [$M_\odot$] & [$M_\odot$] & [$M_\odot$] & [$M_\odot$]\\ \hline
$^{14}$C & 8.73e-10 & 6.33e-08 & 8.84e-10 & 5.84e-08 & 8.62e-10 & 3.51e-07\\
$^{22}$Na & 1.96e-08 & 7.01e-09 & 2.04e-08 & 6.99e-09 & 4.12e-08 & 4.96e-09\\
$^{26}$Al & 5.01e-06 & 3.83e-06 & 5.00e-06 & 3.85e-06 & 5.07e-06 & 1.49e-06\\
$^{32}$Si & 7.27e-12 & 9.32e-11 & 8.88e-12 & 9.69e-11 & 1.12e-11 & 1.72e-08\\
$^{32}$P & 1.43e-08 & 6.92e-08 & 1.43e-08 & 6.73e-08 & 1.50e-08 & 1.13e-06\\
\hline
\end{tabular}
\end{table*}
\newpage
\begin{table*}
\centering
\caption{Nucleosynthesis yields (in solar masses) for select radioactive nuclides of Model M10\_05 with 0.01, 0.1, and 3$\,Z_\odot$.}
\label{app:rad1005_1}
\begin{tabular}{l|cc|cc|cc}
\hline
& \multicolumn{2}{c} {\text{M10\_05\_001}} & \multicolumn{2}{c} {\text{M10\_05\_01}} & \multicolumn{2}{c} {\text{M10\_05\_3}}\\
 & He det & core det & He det & core det & He det & core det\\
 & [$M_\odot$] & [$M_\odot$] & [$M_\odot$] & [$M_\odot$] & [$M_\odot$] & [$M_\odot$]\\ \hline
$^{14}$C & 1.41e-12 & 1.03e-08 & 1.64e-12 & 9.76e-09 & 8.35e-12 & 4.47e-08\\
$^{22}$Na & 3.59e-09 & 7.16e-09 & 3.71e-09 & 8.19e-09 & 7.09e-09 & 6.42e-09\\
$^{26}$Al & 1.11e-07 & 3.88e-06 & 1.11e-07 & 3.89e-06 & 1.24e-07 & 1.64e-06\\
$^{32}$Si & 2.44e-12 & 8.01e-10 & 2.54e-12 & 8.60e-11 & 2.90e-12 & 6.30e-09\\
$^{32}$P & 1.97e-08 & 9.13e-08 & 2.00e-08 & 8.83e-08 & 2.12e-08 & 1.50e-06\\
\hline
\end{tabular}
\end{table*}
\newpage
\begin{table*}
\centering
\caption{Nucleosynthesis yields (in solar masses) for select radioactive nuclides of Model M10\_10 with 0.01, 0.1, and 3$\,Z_\odot$.}
\label{app:rad1010_1}
\begin{tabular}{l|cc|cc|cc}
\hline
& \multicolumn{2}{c} {\text{M10\_10\_001}} & \multicolumn{2}{c} {\text{M10\_10\_01}} & \multicolumn{2}{c} {\text{M10\_10\_3}}\\
 & He det & core det & He det & core det & He det & core det\\
 & [$M_\odot$] & [$M_\odot$] & [$M_\odot$] & [$M_\odot$] & [$M_\odot$] & [$M_\odot$]\\ \hline
$^{14}$C & 3.58e-14 & 7.47e-16 & 7.04e-14 & 7.84e-16 & 1.08e-12 & 2.09e-15\\
$^{22}$Na & 1.25e-08 & 3.21e-13 & 1.22e-08 & 3.20e-13 & 6.93e-09 & 1.65e-10\\
$^{26}$Al & 8.09e-07 & 1.46e-10 & 7.98e-07 & 1.50e-10 & 5.20e-07 & 5.55e-11\\
$^{32}$Si & 4.10e-13 & 2.13e-13 & 5.85e-13 & 2.19e-13 & 5.18e-13 & 6.99e-12\\
$^{32}$P & 3.87e-09 & 1.77e-09 & 3.93e-09 & 1.78e-09 & 4.29e-09 & 1.01e-08\\
\hline
\end{tabular}
\end{table*}
\newpage
\begin{table*}
\centering
\caption{Nucleosynthesis yields (in solar masses) for select radioactive nuclides of Model M11\_05 with 0.01, 0.1, and 3$\,Z_\odot$.}
\label{app:rad1105_1}
\begin{tabular}{l|cc|cc|cc}
\hline
& \multicolumn{2}{c} {\text{M11\_05\_001}} & \multicolumn{2}{c} {\text{M11\_05\_01}} & \multicolumn{2}{c} {\text{M11\_05\_3}}\\
 & He det & core det & He det & core det & He det & core det\\
 & [$M_\odot$] & [$M_\odot$] & [$M_\odot$] & [$M_\odot$] & [$M_\odot$] & [$M_\odot$]\\ \hline
$^{14}$C & 1.02e-12 & 1.39e-16 & 1.15e-12 & 1.46e-16 & 5.25e-12 & 4.72e-16\\
$^{22}$Na & 7.64e-09 & 1.36e-13 & 7.56e-09 & 1.35e-13 & 6.16e-09 & 6.06e-10\\
$^{26}$Al & 2.78e-07 & 3.39e-11 & 2.75e-07 & 3.48e-11 & 2.00e-07 & 1.71e-11\\
$^{32}$Si & 5.17e-13 & 3.42e-14 & 5.32e-13 & 6.53e-14 & 6.65e-13 & 9.46e-13\\
$^{32}$P & 5.12e-09 & 4.12e-10 & 5.18e-09 & 4.19e-10 & 5.77e-09 & 2.17e-09\\
\hline
\end{tabular}
\end{table*}
\newpage

\end{appendix}

\end{document}